\documentclass[useAMS,onecolumn,usegraphicx,usenatbib]{mn2e}
\usepackage{color} 
\usepackage{subfigure}
\newcommand{\LB}{\left[}
\newcommand{\RB}{\right]}
\newcommand{\Lb}{\left(}
\newcommand{\Rb}{\right)}
\newcommand{\be}{\begin{equation}}
\newcommand{\ee}{\end{equation}}
\newcommand{\bea}{\begin{eqnarray}}
\newcommand{\eea}{\end{eqnarray}}
\newcommand{\bmt}{\begin{math}}
\newcommand{\emt}{\end{math}}
\newcommand{\nr}{\rm{NR}}
\newcommand{\Gg}{ {\mathsf{\Gamma \!\! \left( \scriptstyle \frac{3p_1+2}{12} \right) \,\Gamma \!\! \left( \scriptstyle \frac{3p_1+22}{12} \right)}} }

\newcommand{\enorm}{{\mathcal{E}}_{\rm{iso},52}}

\newcommand{\y}{\frac{1-q(2-p_1)}{p_1-1}}
\defcitealias{2001BASI...29..107B}{B01}
\defcitealias{2001ApJ...558L.109D}{DC01}
\defcitealias{2001ApJ...554..667P}{PK01}
\definecolor{gold}{rgb}{0.95,0.1,0.2}

\title[Hard Electron Energy Distribution]{Hard
  Electron Energy Distribution in the Relativistic Shocks of GRB Afterglows}
\author[L. Resmi and D. Bhattacharya]{L. Resmi$^{1,2,3}$\thanks{E-mail:resmi@iap.fr} and D. Bhattacharya$^{3,4}$\thanks{E-mail:dipankar@iucaa.ernet.in}\\
$^{1}$ Institut d'Astrophysique de Paris, Paris 75014, France \\
$^{2}$ Indian Institute of Science, Bangalore 560012, India \\
$^{3}$ Raman Research Institute, Bangalore 560080, India \\
$^{4}$ Inter-University Centre for Astronomy \& Astrophysics, Pune 411007, India}
\begin{document}
\date{}
\maketitle
%
%
\begin{abstract}
Particle acceleration in relativistic shocks is not a very well understood
subject. Owing to that difficulty, radiation spectra from relativistic shocks,
such as those in GRB afterglows, have been often modelled by making
assumptions about the underlying electron distribution. One such assumption is
a relatively soft distribution of the particle energy, which need not be true always, as
 is obvious from observations of several GRB afterglows. In this paper, we describe modifications to the afterglow standard model to accommodate energy spectra which are `hard'.  
We calculate the overall evolution of the synchrotron and compton flux arising
from such a distribution. We also model two afterglows, GRB010222 and GRB020813, under this assumption and estimate the physical parameters. 
\end{abstract}
\section{Introduction}
Relativistic particles accelerated by shocks occupy a predominant place in astrophysical systems. These particles emit synchrotron and compton radiation, which can be observed from radio to gamma-ray bands. Gamma Ray Bursts (GRBs), their afterglows, Supernova Remnants (SNRs), Active Galactic Nuclei (AGNs) and Pulsar Wind Nebulae (PWN) are some of the most important and intriguing candidates which house shock accelerated electron population.  

The details of these electron populations and hence the details of the acceleration
process are inferred from studying the emitted synchrotron and compton
radiation. The accelerated particles are often found to be distributed non-thermally, as a power law in energy characterised by an index p :
\be
N(\gamma_e) = K_e \gamma_e^{-p} ,  (\gamma_m \le \gamma_e < \gamma_{u}),
\ee
where $N(\gamma) d \gamma$ is the number density of electrons in the energy interval 
$\gamma m_e c^2$ and $(\gamma + d\gamma)m_e c^2$.

This non-thermal power law is a natural outcome of the Fermi process
\citep{PhysRev.75.1169}, a standard framework to describe shock acceleration. Several analytical and numerical investigations
have been made \citep{2001MNRAS.328..393A, 2002A&A...394.1141O,
  2004APh....22..323E, 2006PhRvL..97v1104K,2006Ap&SS.tmp..543N} especially for
the Diffusive Shock Acceleration (DSA) mechanism, a variant of the Fermi first
order process, which is expected to operate in collisionless shocks. Most
of the theoretical and numerical studies produce a `single soft' distribution
of the accelerated particles, where the index $p$ is greater than two.

Though there are many observations supporting this prediction, a
non-negligible fraction seems to differ from this. Observations of
some AGNs, GRB Afterglows and PWNs have revealed an underlying `hard' ($p <2$) electron distribution \citep{2001ApJ...560L..49P, 2006MNRAS.371.1441S}. Derivation of expressions for the radiation spectrum from such a distribution requires a different treatment from its `soft' counterpart. 

In this paper, we introduce a modelling platform for afterglow spectral evolution in the presence of a hard electron ($p < 2$) energy distribution. We then present the model of a few afterglows with such a hard spectrum, and derive their physical parameters.  
\section{Hard Electron Energy Spectrum}
The distribution described in equation-1 can be safely assumed to go to
infinity if it is soft, since the role of the higher energy end is negligible
in total number and energy content of the distribution. Hence the equations
which form the basics of the standard afterglow modelling paradigm contain
only $\gamma_m$ and $p$.

However, a hard electron distribution can not be extended upto infinity, and
requires to be terminated with an upper cut off to keep the total energy from
diverging. This upper cut-off, $\gamma_u$, which is determined by the
acceleration mechanism, plays a crucial role in the analytical treatment of $p
<2$ spectra. Since electrons towards the higher energy end dominate in the
share of the total energy content in the distribution, the upper cut-off
appears explicitly in the equations describing the spectral parameters. The
distribution beyond $\gamma_u$ could be a sharp drop, an exponential fall off, or a steeper ($p >2$) powerlaw.

There have been previous studies to incorporate hard electron energy distributions in afterglow modelling. \citeauthor{2001BASI...29..107B} \citeyear{2001BASI...29..107B} (hereafter B01) has used a $\gamma_{u}$ which is a function of the bulk lorentz factor ($\Gamma$) of the shock. The dependence on $\Gamma$ is parametrised by an index $q$.

The time dependence of $\gamma_{m}$ is altered by the introduction of $\gamma_u$. This in turn modifies the spectral evolution. Moreover, a new break frequency corresponding to $\gamma_{u}$ will appear in the spectrum.

\citeauthor{2001ApJ...558L.109D} \citeyear{2001ApJ...558L.109D} (hereafter
DC01) has followed the same approach but by constraining $\gamma_{u}$ (in
their notation, $\gamma_{M}$) to be due to the termination of acceleration
process by energy loss to synchrotron radiation. Their model is a special case of \citetalias{2001BASI...29..107B} with $q = -1/2$. This upper limit $\gamma_{M}$, in typical conditions lie at very high energies.

\citeauthor{2001ApJ...554..667P} \citeyear{2001ApJ...554..667P} (hereafter
PK01) consider two conditions to determine the upper limit of the energy
distribution. (i) The upper limit ($\gamma_{M1}$) results when the
acceleration timescale becomes larger than the timescale for radiative energy
loss (same as DC01), and the corresponding break frequency lies much above the
observation limit. (ii) In the second case, the distribution terminates at an upper cut-off ($\gamma_{M2}$). A steeper 
powerlaw is assumed beyond the cutoff. A constant fraction of the shock
produced thermal energy is assumed to be contained in the electron
distribution, the lower bound of the distribution $\gamma_m$ is assumed to
follow the same evolution as it does in the standard model. The evolution of
$\gamma_{M2}$ results from these two conditions. In the limit, $\gamma_{M2}
\gg \gamma_m$ and $\Gamma \gg 1$, $\gamma_{M2}$ can be obtained analytically
to be proportional to $\Gamma^{-\frac{p-1}{p-2}}$. The second assumption that
$\gamma_m$ follows its standard model behaviour is somewhat inappropriate in
this context, since this behaviour corresponds to a condition where the effect
of $\gamma_{M2}$ is ignorable. In reality, $\gamma_{M2}$ originates in some
physical process which will have its own dependence on $\Gamma$, hence it is more appropriate to parametrise the evolution of $\gamma_{M2}$ as a function of $\Gamma$.

In this paper, we continue the investigation of
\citetalias{2001BASI...29..107B}. The upper cutoff $\gamma_u$ of
\citetalias{2001BASI...29..107B} is identified as an injection break
$\gamma_{i}$, above which the electron distribution steepens to a powerlaw
with index $p_2 > 2$. We leave room for accommodating different processes, by
keeping the parametrisation of $\gamma_i$ to be that of B01. Our results differ from PK01 in having the evolution of $\gamma_m$ and hence of the lightcurve, depending on the nature of the injection break. The flux decay index and the closure relations between the lightcurve decay slope and spectral slope 
also depend on the injection break, essentially the value of $q$, which is characteristic of the mechanism responsible for the upper cut-off.
\section{Modified Electron Distribution and Injection Break}
The {\textit{double slope}} electron energy distribution with slopes $p_1$ and
$p_2$ is represented as,
\be
N(\gamma_e) = \left\{
\begin{array}{ll}
K_e \gamma_e^{-p_1} &  \gamma_m \le \gamma_e < \gamma_i , \\
K_e^{\prime} \gamma_e^{-p_2} &  \gamma_i \le \gamma_e < \infty .
\end{array}
\right.
\ee
Here, $K_e$ is the normalisation constant, which will depend on the number density of the ambient medium $n(r)$ and the bulk lorentz factor $\Gamma$. $K_e^{\prime}$ can be written as, $K_e \gamma_i ^{(p_2-p_1)}$.

We modify the B01 parametrisation of $\gamma_i$ to
\be
\gamma_{i} = \xi(\beta \Gamma)^{q} \label{eq-c2-14} \ \ \ \ \ \ 1 \le \gamma_i
\le \infty .
\ee
in order to accommodate the non-relativistic regime of expansion where $\Gamma \not\gg 1$ and $\beta \not\sim 1$.
Using the standard result that the post shock particle density and energy density are $4 \Gamma n(r)$ and $4 \Gamma (\Gamma -1)n(r) m_{p}c^{2}$ respectively \citep{1998ApJ...497L..17S}, one derives,
\be
K_e = 4 n(r) g_p \; \frac{m_p}{m_e} \; \frac{\epsilon_e}{\xi^{2-p_1}} \; \frac{1}{\beta^{q(2-p_1)}} \; \LB\Gamma -1\RB \; \Gamma^{[1-q(2-p_1)]}  \label{eq-c2-15} 
\ee
\be
\gamma_m = \LB\frac{m_p}{m_e} \frac{\epsilon_e}{\xi^{2-p_1}}f_p\RB^{\frac{1}{(p_1-1)}} \; \LB\frac{1}{\beta^{q(2-p_1)}}\RB^{\frac{1}{(p_1-1)}} \; \LB\Gamma-1\RB^{\frac{1}{p_1-1}} \; \Gamma^{-\frac{q(2-p_1)}{p_1-1}}    \label{eq-c2-16}
\label{pap-eq-3}
\ee
where, $m_p$ and $m_e$ are the proton and electron rest mass respectively. 
The function $g_p = f_p (p_1-1)$ and $f_p = \frac{(2-p_1)(p_2-2)}{(p_1-1)(p_2-p_1)}$.
\subsection{New Spectral Break}
\label{pap-sec-1}
The standard afterglow model has four spectral parameters, the synchrotron
peak frequency, $\nu_m$, the cooling break or the synchrotron cooling frequency, $\nu_c$, corresponding to the
lorentz factor beyond which the electrons cool rapidly, the flux $f_p$ at the peak frequency ($\nu_m$ or $\nu_c$), and the synchrotron
self absorption (SSA) frequency, $\nu_a$, above which the fireball is
optically thin. The radiation spectrum emerging from a double slope electron
distribution will exhibit an additional `` injection break '', corresponding
to the lorentz factor $\gamma_i$. Using the standard expression for
synchrotron frequency $\nu_{\rm{syn}}(\gamma)$ for an electron lorentz factor
$\gamma$, one obtains,
\be
\nu_{i} = {\frac{0.286}{1+z}}{\frac{e}{\pi m_{e}c}} \xi^{2q} \Gamma^{1+2q} \beta^{2q} B,
\ee 
where $B$ is the post-shock magnetic field density, $e$ is the electron
charge, $c$ is the velocity of light and $z$ is the redshift of the burst.
  
Above this frequency the spectral slope steepens to the value corresponding to $p_2$ from that of $p_1$.

%
\section{Spectrum : The source function method} 
Instead of the usual approach of writing flux $f_{\nu}
\propto \nu^{-\delta}$, we use the synchrotron source function along with the
optical depth to obtain the final flux. Therefore, 
\be
f_{\nu} = S_{\nu} [1-\exp(-\tau_{\nu})]
\label{pap-eq-1}
\ee
where $S_{\nu}$ is the synchrotron source function, which has the following functional form:
\be
S_{\nu} = \left\{ \begin{array}{ll}
S_{\nu_p} \Lb \frac{\nu}{\nu_p} \Rb^{2} & \nu < \nu_p \\
S_{\nu_p} \Lb \frac{\nu}{\nu_p} \Rb^{5/2} & \nu > \nu_p
\end{array}
\right.
\label{pap-eq-2}
\ee 
$S_{\nu_p}$ is the source function at peak frequency $ \nu_p$. For slow cooling (ie., $\nu_m < \nu_c$), $\nu_p = \nu_m$ and for fast cooling (ie., $\nu_c < \nu_m$), $\nu_p = \nu_c$. $S_{\nu_p}$ can be calculated as, $S_{\nu_p} = f_p \tau_{\nu_p}$, where $f_p$ is a normalisation constant, that equals the fllux that would have been expected at $\nu_p$ if self absorption were absent. 

The optical depth due to synchrotron process varies as $\nu^{-5/3}$ when $\nu$ is less than $\nu_p$ and $\nu^{-(p+4)/2}$ otherwise. 
Normalising the optical depth to be unity at $\nu = \nu_a$, $\tau_{\nu_p}$, the optical depth at $\nu = \nu_p$ can be written as $\LB \frac{\nu_p}{\nu_a} \RB^{-5/3}$ when $\nu_a < \nu_p$ and $\LB \frac{\nu_p}{\nu_a} \RB^{-(p+4)/2}$ when $\nu_a > \nu_p$. Value of $p$ in the latter expression is $2$ for the fast cooling regime if $\nu_c < \nu < \nu_m$. $p$ is replaced by $p_1$ ($p_1+1$) and $p_2$ ($p_2+1$) in the slow (fast) cooling regime below and above $\nu_i$ respectively.

For a double slope electron energy spectrum undergoing slow cooling, 
\be
\tau_{\nu} = \tau_{\nu_m} \times \left\{ 
\begin{array}{ll}
\Lb \frac{\nu}{\nu_m} \Rb^{-5/3} & \nu < \nu_m  \\
\Lb \frac{\nu}{\nu_m} \Rb^{-(p_1+4)/2} & \nu_m < \nu < (\nu_c,\nu_i) \\
\Lb \frac{\nu_c}{\nu_m} \Rb^{-(p_1+4)/2} \Lb \frac{\nu}{\nu_c} \Rb^{-(p_1+5)/2} & \nu_m < \nu_c < \nu < \nu_i \\
\Lb \frac{\nu_i}{\nu_m} \Rb^{-(p_1+4)/2} \Lb \frac{\nu}{\nu_i} \Rb^{-(p_2+4)/2} & \nu_m < \nu_i < \nu < \nu_c \\
\nu_m^{(p_1+4)/2} \, \nu_c^{1/2} \, \nu_i^{(p_2-p_1)/2} \, \nu^{-(p_2+5)/2} & (\nu_i,\nu_c) < \nu
\end{array}
\right. 
\ee

For fast cooling,

\be
\tau_{\nu} = \tau_{\nu_c} \times \left\{ 
\begin{array}{ll}
\Lb \frac{\nu}{\nu_c} \Rb^{-5/3} & \nu < \nu_c  \\
\Lb \frac{\nu}{\nu_c} \Rb^{-3} & \nu_c < \nu < \nu_m\\
\Lb \frac{\nu_m}{\nu_c} \Rb^{-3} \Lb \frac{\nu}{\nu_m} \Rb^{-(p_1+5)/2} & \nu_c < \nu_m < \nu < \nu_i \\
\Lb \frac{\nu_m}{\nu_c} \Rb^{-3} \Lb \frac{\nu_i}{\nu_m} \Rb^{-(p_1+5)/2} \Lb \frac{\nu}{\nu_i} \Rb^{-(p_2+5)/2} & \nu_c < \nu_m < \nu_i < \nu
\end{array}
\right. 
\ee

These expressions, along with equation-\ref{pap-eq-2} are substituted in
equation-\ref{pap-eq-1} to obtain the final flux, which at a given time, is
a function of the five spectral parameters ($\nu_m$, $\nu_a$, $\nu_c$, $\nu_i$ and
$f_p$). 

To estimate these parameters, we first evaluate $\Gamma(r)$ and
$r(t)$. For that, we use the expressions given by \citet{2000ApJ...543...90H},
after correcting for redshift, which accommodates a smooth transition from an
initial ultra-relativistic to the final non-relativistic regime of the fireball. 
Time evolution of the half opening angle ($\theta_j$) depends on the lateral
velocity of the jet in its comoving frame, which essentially is the sound
velocity of the post-shock medium. The half opening angle varies as, $\frac{d\theta_{j}}{dr} = \frac{1}{\beta \Gamma}\left[\frac{c_{s}}{c}\right]$,
where $c_s$ is the velocity of sound in the downstream medium. $c_s$ is
usually assumed to be constant throughout the evolution of the shock, but this is
not a very accurate assumption. Initially, when the downstream plasma is
ultra-relativistic, the thermal velocity will be $c/\sqrt{3}$, but as the
ejecta becomes non-relativistic, the velocity approaches
$\sqrt{\frac{k_BT}{m_p}}$, where $m_p$ is rest mass of the proton. We calculate $c_s$ as a function of $\Gamma$,
adopting the method followed by \citet{1939isss.book.....C}. This gives us 
\be
\left[ \frac{c_{s}}{c}\right]^{2} = \frac{k_B T}{m_p c^2} \frac{1}{\Gamma}
\label{eq-c2-09}
\ee
We have used equation-A3 (Appendix-I) to obtain temperature in terms of $\Gamma$. More details of the calculation is given in the Appendix.
The comoving magnetic field density $B$ is given as $\LB 8\pi 
\epsilon_{B} \frac{(\Gamma-1)m}{V_{\rm{co}}} \RB^{\frac{1}{2}}c $, where
$\epsilon_{B}$ is the fraction of thermal energy in the magnetic field, $m$ is the total swept up mass,
  $V_{\rm{co}}$ is the volume of the downstream plasma in the comoving frame,
  which can be calculated as $\Omega r^{2} \Delta^{\prime}$ where $\Omega$ is the solid angle and 
  $\Delta^{\prime}$ is the comoving shell thickness.

We calculate $f_p$ using the expression (equation-25) given by
\citet{1999ApJ...523..177W}. $\nu_m$ and $\nu_c$ are calculated using the
expression described in section \ref{pap-sec-1}, by replacing $\gamma_i$ with
$\gamma_m$ (equation-\ref{pap-eq-3}) and $\gamma_c$ ($ = 6 \pi m_e c/(\sigma_T \Gamma B^2
t)$). $\nu_a$ is the frequency at which the synchrotron optical depth in the
comoving frame ($\alpha_{\nu^{\prime}}^{\prime} \Delta^{\prime}$, where
$\alpha_{\nu}$ is the absorption coefficient calculated following the method
given by \citet{1979rpa..book.....R}) equals unity.
 
For various values of $q$, the evolution of the spectral breaks as a function
of time is plotted in figure 1 and the lightcurves are displayed in figure 2. The difference of evolution introduced by $q$ is apparent in these figures.

\section{Dynamics : Limiting Cases}

To obtain the overall dynamics of the fireball, we adopt the method presented
by \citet{2000ApJ...543...90H} which accomodates a smooth transition from the
initial ultra-relativistic to the final non-relativistic phase. 

However, analytical solutions for $\Gamma(r)$ are possible in extreme
cases. The adiabatic ($\epsilon = 0$) ultra-relativistic regime ($\Gamma \gg
1, \beta \sim 1$) is encountered most commonly in afterglow observations. At
late times, ($t > t_{\rm{NR}}$, the fireball becomes non-relativistic. This
phase is same as that of the well studied supernova remnants. 

\subsection{Ultra-relativistic Limit}
In this limit, the expressions for $\Gamma(r)$ and $r(t)$ of
\citet{2000ApJ...543...90H} can be approximated to \\ $\sqrt{(3-s) E_0/(\Omega c^2}) \, \,(\rho_0 r_0^3)^{-1/2} \, \, (r/r_0)^{(s-3)/2}$ and $((4-s)(3-s)2ctE_0 / ((1+z)\Omega c^2 \rho_0r_0^s)) ^{\frac{1}{4-s}}$ respectively, where $\rho(r)$, the ambient medium mass density profile is parametrised as $\rho_0 \, (r/r_0)^{-s}$. 
The expressions for spectral parameters we obtained for this phase, are listed below. We consider two types of ambient media, (i) a constant density around the progenitor star ($n(r) =n, s=0$) and (ii) a stellar-wind blown stratified density profile ($s=2$, with a normalisation $\rho_0 = 5 \times 10^9 A_{\star}$ and $r_0 = 10^{10}$cm).
%
\be
f_p (mJy)= \left\{
\begin{array}{lr}
210.45\;\phi_{p_1}\; \frac{1+z}{d_{L,{\rm{Gpc}}}^2} \sqrt{\frac{c_s}{c}}\; \sqrt{\epsilon_B\,n}\; \enorm & (s=0)   \\
 1021.5\, \frac{\phi_{p_1} (1+z)}{ d_{L,{\rm{Gpc}}}^2} \sqrt{\enorm \epsilon_B}  A_{\star} \LB \frac{t_d}{(1+z)} \RB^{-1/2} & (s=2) \\
\end{array}
\right.
\ee
\be
\nu_m (Hz) = \left\{
\begin{array}{lr}
1.87 \times 10^{7}\; (17.14)^{\y}\; \LB \frac{m_p}{m_e} \: f_p \RB^{\frac{2}{p_1-1}} \;  \sqrt{\frac{c_s}{c}} \; \frac{x_{p_1}}{1+z}\; \sqrt{\epsilon_B n} \; \epsilon_e^{\frac{2}{p_1-1}}  &  \nonumber \\
  \xi^{-2\frac{2-p_1}{p_1-1}} \;  \LB \frac{\enorm}{n}\RB^{\frac{p_1+qp_1-2q}{4(p_1-1)}} \; \frac{t_d}{(1+z)}^{-\frac{3(p_1+qp_1-2q)}{4 (p_1-1)}}  & (s=0) \\
%
%
5.77 \times 10^7 \; (13.1)^y \; \sqrt{\frac{c_s}{c}} \, \frac{x_{p_1}}{1+z} \, \LB \frac{m_p}{m_e} f_{p_1}\RB^{\frac{2}{p_1-1}} \; \enorm^{y/2} \; A_{\star}^{(1-y)/2} & \nonumber \\
\sqrt{\epsilon_B}  \; \epsilon_e^{2/(p_1-1)} \;  \xi^{\frac{-2}{(2-p_1)(p_1-1)}} \;  \LB \frac{t_d}{(1+z)}\RB^{\frac{-(2+y)}{2}}  & (s=2) \\
\end{array}
\right.
\ee
where $y= \frac{1-q(2-p_1)}{p_1-1}$, $\phi_p$ and $x_p$ are functions of $p$ \citep{1999ApJ...523..177W}.
%
%
%
%
%
\be
\nu_c (Hz) = \left\{
\begin{array}{lr}
5.84 \times 10^{13} \;  \LB{\frac{c_s}{c}}\RB^3 \; \enorm^{-1/2} \; n^{-1} \; \epsilon_B^{-3/2}\; [t_d (1+z)]^{-1/2} & (s=0) \\
7.6 \times 10^{11} \; \frac{1}{(1+z)^3} ;\ {\frac{c_s}{c}}^{-3/2} \: {\epsilon_B}^{-3/2} \; A_{\star}^{-2} \; \enorm^{1/2} \; \LB\frac{t_d}{(1+z)}\RB^{1/2} & (s=2) \\
\end{array}
\right.
\ee
%
%
%
%
\be
\nu_i (Hz) = \left\{
\begin{array}{lr}
1.3 \times 10^6 \; \frac{(4.14)^{1+2q}}{1+z}\; \sqrt{\frac{c_s}{c}}\; \xi^2\;\sqrt{\epsilon_B\,n} \; \LB\frac{\enorm}{n}\RB^{\frac{1}{4}(1+q)} \; \LB\frac{t_d}{(1+z)}\RB^{\frac{-3}{4}(1+q)} & (s=0) \\
1.65 \times 10^7 \; \xi^2 (3.62)^q \sqrt{\frac{cs}{c}} \enorm^{q/2}A_{\star}^{(1+q)/2} \epsilon_B^{1/2} \LB\frac{t_d}{(1+z)}\RB^{\frac{-1}{2}(2+q)} & (s=2)
\end{array}
\right.
\ee
%
%
In the slow cooling regime,
\be
\nu_a ({\rm{Hz}}) = \left\{
\begin{array}{l}
(2.61 \times 10^{19})^{\frac{p_1}{4+p_1}} \; (8.38 \times 10^{19})^{\frac{2}{4+p_1}} \; (8.2 \times 10^{-7})^{\frac{2(p_1-1)}{4+p_1}} \; (0.88)^{\frac{2+p_1}{4+p_1}} \; (3)^{\frac{p_1+1}{p_1+4}} \\
(4.14)^{\frac{2(2-2q+p_1q)}{4+p_1}} \; (0.64)^{\frac{2+p_1}{4+p_1}} \;  \LB\sqrt{\frac{c}{c_s}}\RB^{\frac{p_1+2}{p_1+4}} \\ 
(2.3 \times 10^{-10})^{\frac{p_1}{p_1+4}} \LB \Gg \RB^{2/(p_1+4)} \; \LB 1.87 \times 10^{-12} g_p \sqrt{3}\frac{c_s}{c}\RB^{\frac{2}{p_1+4}}  \\
\enorm^{\frac{p_1+6+qp_1-2q}{4(p_1+4)}} \; \epsilon_B^{\frac{p_1+2}{2(p_1+4)}} \; \epsilon_e^{2/(p_1+4)} n^{\frac{p_1+6-qp_1+2q}{4(p_1+4)}}\; \xi^{-2\frac{2-p_1}{p_1+4}}  \\
\LB {\frac{t_d}{1+z}} \RB ^{\frac{-10-3p_1-3qp_1+6q}{4(p_1+4)}}  \: \: \: \: (s=0 \: \: \: \nu_a > \nu_m)\\
\\
2.61 \times 10^{14} \;  \; (4.14)^{\frac{10q+3p_1-5qp_1-8}{5(p_1-1)}} \; \LB \frac{c_s}{c}\RB^{23/40} \; \LB f_p \frac{m_p}{m_e} \RB^{-\frac{3p_1+2}{5(p_1-1)}} \\
\LB \frac{(4-p_1^2)(p_2-2)}{(p_1+2/3)(p_2-p_1)} \RB^{3/5} \;  \epsilon_B^{1/5} \; \epsilon_e^{\frac{-1}{(p_1-1)}} \xi^{\frac{2-p_1}{p_1-1}}  \; \enorm^{\frac{18-13p_1-10q+5p_1q}{40(p_1-1)}} \; n^{\frac{14-19p_1+10q-5p_1q}{40(p_1-1)}} \;      \\
\LB \frac{t_d}{1+z} \RB ^{\frac{3(p_1-2)(q-1)}{8(p_1-1)}}  \: \: \: \: (s=0 \: \: \: \nu_a < \nu_m)\\
\end{array}
\right.
\ee
where $\mathsf{\Gamma}$ denotes the Gamma function.
\be
\nu_a({\rm{Hz}}) = \left\{
\begin{array}{l}
(3.62)^{y_2} \; (3.42 \times 10^{53})^{\frac{1}{(p_1+4)}} \; ( 2.4 \times 10^7 )^{\frac{p_1}{(p_1+4)}} \; \LB g_p \frac{\epsilon_e}{\xi^{2-p_1}} \RB^{\frac{2}{(p_1+4)}}  \nonumber \\
\LB \sqrt{\frac{c_s}{c}} \sqrt{\epsilon_B}\RB^{\frac{p_1+2}{p_1+4}} \; \Gg \; \LB \frac{c_s}{c} \RB^{\frac{2}{p_1+4}}  \nonumber \\
\enorm^{\frac{qp_1 -2q}{2(p_1+4)}} \; A_{\star}^{\frac{1.+2p_1+4q-2p_1q}{p_1+4}} \; \LB \frac{t_d}{(1+z)} \RB^{\frac{2q-2p_1-qp_1-8}{2(p_1+4)}} \: \: \: \: (s=2  \: \: \: \nu_a > \nu_m)  \\
6.16 \times 10^{14} \; 3.62^{y_3} \; \LB \frac{p_1+2}{p_1+2/3}\RB^{3/2} \; \LB \frac{c_s}{c} \RB^{23/40} \; \LB \frac{m_p}{m_e} f_p\RB^{-\frac{2+3p_1}{5(p_1-1)}} \nonumber \\
g_p^{3/5} \epsilon_B^{1/5} \; \enorm^{\frac{2}{5} + \frac{(p_1-2) (1-q)}{4(p_1-1)}} \; A_{\star}^{\frac{6}{5} - \frac{(p_1-2)(1-q)}{4(p_1-1)}}\; \LB \frac{\epsilon_e}{\xi^{2-p_1}} \RB^{\frac{3}{5} - \frac{2+3p_1}{5(p_1-1)}}\nonumber \\
\LB \frac{t_d}{(1+z)} \RB^{\frac{7p_1-2-10q+5p_1q}{20(p_1-1)}} \: \: \: \:(s=2  \: \: \: \nu_a < \nu_m)
\end{array}
\right.
\ee
where $y_2 = \frac{p_1+6-4q+2qp_1}{p_1+4}$ and $y_3 = \frac{(p_1-2)(1-q)}{p_1-1}$

%
%
%
%
In the fast cooling regime,
\be
\nu_a({\rm{Hz}}) = \left\{
\begin{array}{lr}
1.96 \times 10^{11} \; (p_1-1)^{1/3} \, \LB \frac{c_s}{c} \RB^{9/2} \, (1+z)^{2/3} \; \enorm^{1/6} \; n_0^{1/6} \; \LB \frac{t_d}{1+z}\RB^{-1/2}   & \nu_c < \nu_a < \nu_m  \:\:  \: \: (s=0) \\
1.83 \times 10^{10} \; (p_1-1)^{3/5} \, \LB \frac{c_s}{c} \RB^{57/10} \, (1+z)^2 \; \enorm^{7/10} \; n_0^{11/10} \; \epsilon_B^{6/5} \; \LB \frac{t_d}{1+z}\RB^{-1/2} & \nu_a < \nu_c \: \: \: (s=0)  \\
1.44 \times 10^{9} \; (p_1-1)^{1/3} \, A_{\star}^{1/3} \;  \LB t_d(1+z) \RB^{-2/3} & \nu_c < \nu_a < \nu_m  \:\:  \: \: (s=2) \\
8.48 \times 10^7 \; (p_1-1)^{3/5} \;  \LB \frac{c_s}{c} \RB^{6/5} \; \enorm^{-2/5} \; A_{\star}^{11/5} \;\epsilon_B^{6/5} \; \LB\frac{t_d}{1+z}\RB^{-8/5} \;  & \nu_a < \nu_c \: \: \: (s=2)  
\end{array}
\right.
\ee

\subsubsection{$\alpha$-$\delta$ closure relations}
The $\alpha$-$\delta$ closure relations for a general value of $q$ valid in the slow cooling phase of the ultra-relativistic approximation are the following:
\be
\alpha=\left\{
\begin{array}{ll}
\frac{3}{8} \, [(q-1) - 2 \delta (q+1)] & \nu_m < \nu < \nu_c, \; \; t<t_j\\
\frac{1}{4} \, [(3q-1)- 3 \delta(q+1)] & \nu > \nu_c, \; \; t<t_j \\
\frac{1}{2} \, [q - 3 - 2 \delta (q+1)] & \nu_m < \nu < \nu_c, \; \; t>t_j \\
(q-1) - \delta(q+1) & \nu > \nu_c, \; \; t > t_j
\end{array}
\right. 
\label{eq-closure}
\ee
In figure 3., we display the above closure relations. The
$q=1$ plot can be considered as a reference to the standard model, as it
recovers the usual slopes. The dependence $\alpha$ has on $q$ has to be kept
in mind while inferring the value of $p$ from the lightcurves. Temporal decay
indices calculated for the ultra relativistic limit are listed in table 1 and lightcurve decay indices
are listed in table 2 (slow cooling) and in table 3 (fast cooling).
\subsection{Non-relativistic Limit}
In the non-relativistic limit, at $t = t_{\rm{NR}}$, the lorentz factor 
is $\sim 1$ and $\beta \ll 1$. The fireball by this time would have undergone a considerable lateral spread and the geometry may be approximated to be spherical. The solid angle $\Omega$ may now be set to $4 \pi$.

\subsubsection{Dynamics}
The evolution of the radius $r$ is calculated as, 
\be
r = \zeta(\hat{\gamma}) \, \LB E_0 \, \frac{r_0^s}{\rho_0} \RB^{1/(5-s)} \, t^{2/(5-s)}
\ee
where $E_0$ is the energy in the explosion and $\hat{\gamma}$ is the ratio of specific heats for the plasma. One could assume $\zeta(\hat{\gamma})$ to be $1.05$ for a constant density ambient medium and $0.65$ for a stellar-wind blown medium \citep{2004ApJ...612..966B}.

%
%
%
\subsubsection{Electron energy spectrum}
The thermal energy density in the shock downstream is estimated as,
\be
u_{\rm{th}} = \frac{9c^2 \rho_0}{8} \; \beta^2\; (r/r_0)^{-s}
\ee
where $\beta$ is $\frac{1}{c}\frac{dr}{dt}$.
The expressions for electron number and energy will give, respectively,
\be
\frac{K_e}{(p_1-1)\gamma_m^{p_1-1}} = 4 \rho_0/m_p \; (r/r_0)^{-s} \label{eq-c2-23}
\ee
\be
\frac{K_e m_e}{g_p}\; \gamma_i^{(2-p_1)} = \frac{9c^2\rho_0}{2}\epsilon_e \; \beta^2 \; (r/r_0)^{-s}  \label{eq-c2-24}
\ee
Solving eq. \ref{eq-c2-24} and eq. \ref{eq-c2-23}, one obtains the expressions for $K_e$ and $\gamma_m$:
\be
K_e =\frac{9}{2} g_p \frac{\rho_0}{m_e} \; \frac{\epsilon_e}{\xi^{2-p_1}} \; (r/r_0)^{-s} \; \beta^{2-q(2-p_1)}
\ee
\be
\gamma_m = \LB \frac{9}{8} f_p \frac{m_p}{m_e} \frac{\epsilon_e}{\xi^{2-p_1}}  \RB^{1/(p_1-1)} \; \beta^{\frac{2-q(2-p_1)}{p_1-1}}
\ee
\subsubsection{Spectral Parameters}
The magnetic field energy density is assumed, as usual, to be a fraction $\epsilon_B$ times the thermal energy density. ie.,
\be
B = \sqrt{9 \pi \epsilon_B \beta^2 c^2 \rho(r)}
\ee

We calculate the four spectral breaks, $\nu_a$, $\nu_m$, $\nu_c$, and $\nu_i$ and the peak flux $f_p$ from:

\begin{eqnarray}
f_p  & = & \frac{2.94 \times 10^{-21}}{d_L^2\,a} \; 0.053^{(p_1-1)} \; \Gg \; r^3 \frac{K_e \; B}{\gamma_m^{p_1-1} }  \\
\nu_m & = & 2.8 \times 10^{6} \; \frac{x_p}{(1+z)} \; B \; \LB 2065.7 f_p \frac{\epsilon_e}{\xi^{(2-p_1)}}\RB ^{2/(p_1-1)} \; \beta^{\frac{2[2-q(2-p_1)]}{(p_1-1)}} \\
\nu_c & = & 4.81 \times 10^{23} \: \frac{1}{B^2} \LB \frac{t}{t_{\nr}} \RB^{-2} \; \LB \frac{t_{\nr}}{1+z} \RB^{-2}  \\
\nu_i & = & 8.0 \times 10^{5}\: \frac{1}{(1+z)} \: B \xi^2 \beta^{2q}
\end{eqnarray}
\be
\nu_a =
\left\{
\begin{array}{l}
4.72 \LB \frac{p_1+2}{p_1+2/3} \: K_e \RB ^{3/5} \: \gamma_m^{-(3p_1+2)/5} \; \; r^{3/5} B^{2/5} \: \: \:  \mbox{(for $\nu_a > \nu_m$)}   \\
 \\
(6.72 \times 10^{-13})^{\frac{p_1-1}{p_1+4}} \: (1.25 \times 10^{19})^{\frac{p_1}{p_1+4}} \; (7 \times 10^{-5})^{\frac{1}{p_1+4}}  \\
\LB \Gg \RB^{\frac{2}{p_1+4}} \; B^{\frac{p_1+2}{p_1+4}} \; \; \LB \frac{rK_e}{a} \RB ^{\frac{2}{p_1+4}} \: \: \: \: \: \mbox{(for $\nu_a < \nu_m$)} \\
\end{array}
\right.
\ee
where $a$ is a numerical factor, describing the thickness of the shock in terms of $r$ as $\Delta = r/a$
\section{Synchrotron Self Compton Emission}
The contribution to the total flux from synchrotron photons which are compton
scattered by the non-thermal relativistic electrons themselves, can be significant towards higher energies.
 
We calculate this compton component following the method adopted by
\citet{2001ApJ...548..787S}. Following this work, the approximate ratio of
inverse compton (IC) to synchrotron luminosities may be estimated as follows
(for a uniform density ambient medium and the slow cooling regime). 
 
The IC spectrum is characterised by four break frequencies : $\nu_m^{\rm{IC}}
=  2 \gamma_m^2 \nu_m^{\rm{syn}}$,  $\nu_c^{\rm{IC}}  =  2 \gamma_c^2
\nu_c^{\rm{syn}}$,  $\nu_i^{\rm{IC}}  =  2 \gamma_i^2 \nu_i^{\rm{syn}}$,
 $\nu_a^{\rm{IC}}  =  2 \gamma_m^2 \nu_a^{\rm{syn}}$ and a flux normalisation $f_p^{\rm{IC}}  =  f_p^{\rm{syn}} \sigma_T \,  n \,  r$

For $\nu_m^{\rm{syn}} \le \nu_i^{\rm{syn}} \le \nu_c^{\rm{syn}}$, the energy
emitted by the compton process peaks at $\nu_c^{\rm{IC}}$ and that by the synchrotron process will peak at $\nu_c^{\rm{syn}}$.
\bea
x & \equiv \frac{L^{\rm{IC}}}{L^{\rm{syn}}} & \approx \frac{ \nu_c^{\rm{IC}} \;
  f_{\nu_c}^{\rm{IC}}} { \nu_c^{\rm{syn}} \; f_{\nu_c}^{\rm{syn}}} \\ \nonumber
& & = 700 {\mathcal{R}}_{-7} \;  \gamma_{c,7}^2 \;  \LB
\frac{\gamma_{m,500}}{\gamma_{i,5}} \RB^{(p_1-1)_{0.5}} \;  \LB
\frac{\gamma_{i,5}}{\gamma_{c,7}} \RB^{(p_2-1)_{1.5}}  
\eea

For $\nu_m^{\rm{syn}} \le \nu_c^{\rm{syn}} \le \nu_i^{\rm{syn}}$, compton energy peaks at $\nu_i^{\rm{IC}}$ and synchrotron energy peaks at $\nu_i^{\rm{syn}}$  
\bea
x & \approx & \frac{ \nu_i^{\rm{IC}}  \; f_{\nu_i}^{\rm{IC}}}{ \nu_i^{\rm{syn}} \;  f_{\nu_i}^{\rm{syn}}} \\ \nonumber
& \approx & 700 {\mathcal{R}}_{-7} \;  \gamma_{i,7} \; \gamma_{c,5} \; \LB \frac{\gamma_{m,500}}{\gamma_{i,7}} \RB^{(p_1-1)_{0.5}}
\eea

where $\gamma_{e,n} = \gamma_e/10^n$, $\gamma_{m,500} = \gamma_m/500$, ${\mathcal{R}}_{-7} = \frac{f_p^{\rm{IC}}/f_p^{\rm{syn}}}{10^{-7}}$ and $(p-1)_{f} = (p-1)/f$.
In either case, the compton power peaks at very high frequencies, ($\sim
10^{21}$~Hz $\frac{B}{0.1 {\rm{G}}} \frac{\Gamma}{100}$) for a hard electron
spectrum. Hence the contribution of synchrotron self compton emission becomes significant only at frequencies above hard x-rays.

As a next step, we estimate the IC flux from a numerical integration over the
photon and the electron spectra.
To do so we use the expression given by \citet{2001ApJ...548..787S} for the inverse compton flux due to the modified electron distribution, and the synchrotron radiation spectrum $f_{\nu}^{\rm{syn}}$ generated by this electron energy spectrum,

\be
f_{\nu}^{\rm{IC}} = r \sigma_T \int_{\gamma_m}^{\infty} d\gamma \, N(\gamma) \int_{0}^{x_0} dx f_{\nu}^{\rm{syn}}(x)
\ee
where $x_0 \sim 0.5$

The synchrotron and the compton fluxes obtained from the above calculation are displayed in figure 4.

\section{Modelling Shallow Evolution}
A new parameter $q$ is required for modelling afterglow evolution based on hard electron energy spectrum. This index parametrises the evolution of the upper cut-off of the electron spectrum (see equation-3). The value of $q$ is determined by the acceleration process operating in the relativistic shocks. The present understanding about this from theoretical or numerical calculations is not exhaustive. 

The termination of the acceleration process due to synchrotron radiation losses leads to $\gamma_i$ being inversely proportional to the square-root of the bulk lorentz factor ($q=-0.5$) \citep{1999MNRAS.305L...6G,2006ApJ...651..328L}. However, the slowest post jet break decay in this case tends to $1.75$ as $p_1$ tends to its minimum possible value of $1$ (in the limit $1 \le p_1 \le 2$). This is noticed by DC01 also, who have tried to model GRB010222 using a hard electron energy spectrum. They have used this fact to rule out the presence of a hard electron energy distribution in this afterglow. None of the afterglows we model in this paper however display post jet break decays steeper than $1.75$, which rules out the possibility of their electron distribution be terminated by synchrotron losses. 

$q=1$ is applicable to the lower cutoff of fermi process ($\gamma_i = \frac{m_p}{m_e} \, \Gamma$), below which a pre-acceleration mechanism producing a flat electron spectrum may operate \citep{2001AIPC..558..392A}. The presence of such an upper cut-off is observed in some of the Active Galactic Nuclei \citep{1989MNRAS.239..401L,2003ApJ...597..851K,2007ApJ...662..213S} and Pulsar Wind Nebulae \citep{1992ApJ...390..454H}. Moreover, $q=1$ also provides scalings that would have been obtained in the standard fireball model without references to $\gamma_i$. Good fits could be obtained with a $q$ of $1$ for all three afterglows we study \citep{2004ASPC..312..411B,2005BASI...33..487M}, however, the value of $\xi$ we inferred from these fits are far higher than $m_p/m_e$. 

Another interesting value of $q$ is $-1.0$, though any mechanism producing such an upper cut-off proportional to the inverse of the bulk lorentz factor is not discussed in the literature to the best of our knowledge. $q =-1$ provides $\alpha_1$ of $0.75$ and $\alpha_2$ of $2.0$, independent of the value $p$ assumes, as is obvious from equation-\ref{eq-closure} since $\delta$ is always multiplied by $(q+1)$, which in this case vanishes.
It is interesting that these $\alpha$~s correspond to $p > 2$ scaling relations if applied to a $p$ of $2.$ 

For GRB afterglows, it is not often very easy to infer the value of $p$
unambiguously. The spectral index estimated from observations in the optical
bands is a composite of the unknown host galaxy extinction and the intrinsic
spectral index, $\delta$. The X-ray spectrum is not affected by dust
extinction but is modified by photoelectric absorption at lower energies. This
makes the x-ray spectral index to be a function of the unknown gas column
density along the line of sight. Also, due to the low count rate, it is often
difficult to bin the spectrum and get the value of $\delta$ accurately. A
third method is to measure the flux decay index past the jet break in optical
and in x-ray wavelengths and assume it to be $p$, as predicted by the standard
afterglow model. Though it suffers from complexities in the modelling of the
fireball dynamics, this method is
largely followed and trusted. However, the spectral index derived should be
consistent with the closure relations between the temporal decay index,
$\alpha$ and the spectral index, $\delta$ in various bands.

Recently several studies have suggested the possibility that the electron energy index, inferred by some of the above methods, falls below $2$. Out of the $16$ well observed pre-{\textit{Swift}} afterglows studied by \citet{2006ApJ...637..889Z}, $\alpha_2$ of five afterglows fall below $2$. \citet{2006MNRAS.371.1441S} along with blazars and PWNs, study a sample of well monitored X-ray afterglows observed by {\textit{BeppoSAX}} and {\textit{Swift}}. The inferred values of $p$ fall below $2$ for eight of them (See figure 5 of \citet{2006MNRAS.371.1441S}). Early evolution of several x-ray afterglows monitored by {\textit{Swift}} have  shown an unprecedented `flat' evolution \citep{2006ApJ...642..389N}. Though not all of them may have an intrinsic flat electron energy spectrum (some could show shallow decay due to prolonged energy injection from the central engine), some are well within the expectations of hard spectrum models.
In some of the {\textit{Swift}} x-ray lightcurves (for example, GRB050820, GRB051109A, GRB061024), the normal decay phase, which follows the shallow phase, has $\alpha$ values expected from an underlying hard electron energy spectrum \citep{2008ApJ...675..528L}.

In the following section, we model three pre-{\textit{Swift}} afterglows, with
rich multiband data set, showing evidence of an underlying hard electron
energy spectrum. We consider $q$ as a fit parameter and use a range of $-2. < q < +2$ while searching for the best fit. 

\subsection{GRB010222}
GRB 010222 \citep{2001GCN...959....1P}, at a redshift of 1.477
\citep{2001ApJ...554L.155J, 2002ApJ...578..818M} was one of the first
afterglows seen with hard electron spectrum and it initiated theoretical work 
in that direction (B01, DC01). 

The optical afterglow evolution was initially shallow ($\alpha_1 \sim
0.6$) and it steepened to an $\alpha_2$ of $1.3$ -- $1.4$ around $\sim 0.5$~day
\citep{2001BASI...29...91S,2001ApJ...563..592S}. Around the same time the
x-ray lightcurve also steepened from $\alpha_1 \sim 0.6$ to $\alpha_2 \sim 1.3$
\citep{2001ApJ...559..710I}. Assuming this early achromatic break to be due to the lateral expansion of the jet, a hard electron distribution is required to explain the evolution past this 
break.
The spectral index, $\delta_o$, within the optical band was found to be $0.89 \pm
0.03$ after correcting for Galactic extinction
\citep{2002ApJ...578..818M}. The x-ray spectral index ($\delta_x$) depends on the assumed value of neutral hydrogen column density of the
host galaxy. \citep{2001ApJ...559..710I, 2002ApJ...579L..59B}, however it
falls in the range of $0.7$ -- $0.9$. 

Our model with $p_1 \sim 1.5$ and $q \sim 1.3$ reproduces the
observed lightcurve decay indices before and after the jet break. We assume
$\nu_c$ to be below both optical and x-ray bands at $\sim 0.5$~day and $\nu_i$ to be above the x-ray bands. Along
with the extinction in the host galaxy ($E_{B-V} = 0.03$; starburst type
extinction law by \citet{1997AJ....113..162C}) this reproduces the observed optical and x-ray spectrum. 

A model with $q$ of $1.0$ and $\nu_i$ in x-ray bands reproduces the data fairly well \citep{2004ASPC..312..411B} and also explains the spectral steepening seen towards the x-ray band (the x-ray spectral index derived by \citet{2001ApJ...559..710I} using the 
{\textit{Beppo-SAX}} data , is steeper than that in 
the optical bands). However, 
our best fit is obtained when $q$ is $1.3$, not when it is unity. A higher $q$ requires a steeper $p_1$ to reproduce the lightcurves decay indices as $\delta_1$ and $\delta_2$ decrease as $q$ increases. The best fit with $q=1.3$ (figure 5) requires that $\nu_i >\nu_x$. 

We calculated the inverse compton
emission for these parameters, and found that it is negligible at the x-ray
frequencies. We obtain a peak flux $f_p$ of $1.04$~mJy and the peak frequency
$\nu_m$ of $\sim 200$~GHz, at the time of the break. From these fit
parameters, we infer an isotropic equivalent energy of $5.9 \times 10^{52}
n_0^{1/5}$~erg, a jet opening angle of $2.1^{\circ} n_0^{1/10}$, and a total
energy of $3.6 \times 10^{49} n_0^{2/5}$~ergs. An upper
limit of $10^5$ is estimated for $\xi$. The best fit model along with the observations are displayed in figure 5. The spectral parameters and physical parameters are listed in table 4 and table 5 respectively.

We note that a model assuming continuous energy injection by
\citet{2002ApJ...579L..59B} can also reproduce the observed evolution of this
afterglow. Another explanation for the achromatic break observed around $\sim
0.5$~day is the non-relativistic transition of the fireball \citep{2001A&A...374..382M}, but
such an early non-relativistic transition would require a very high ambient
medium density ($n \sim 10^{6}$~atom/cc for the observed fluence of this burst) which would
have suppressed the radio flux to nano-jansky levels.

\subsection{GRB020813}
GRB020813 was detected by HETE-II \citep{2002GCN..1471....1V} at a redshift of
1.26 \citep{2002GCN..1475....1P}. The optical afterglow of this burst, like
GRB010222, exhibited a shallow decay and an early break ($\alpha_1 \sim 0.8$ ,
$t_b \sim 0.5$~day in optical \citep{2003A&A...404L...5C}). The x-ray
observations started after the optical break, the lightcurve exhibited a single power law decay consistent with the post break optical decay ($\alpha_o \sim 1.4$ \citep{2003A&A...404L...5C}, $\alpha_x \sim 1.4$ \citep{2003ApJ...597.1010B}). The optical photometric spectral index, corrected for Galactic absorption was $\sim 0.9$ \citep{2003A&A...404L...5C} and the x-ray spectral index was $\sim 1.0$ \citep{2003ApJ...597.1010B} with no absorption column in excess of the Galactic value of $7.5 \times 10^{20}$~cm$^{-2}$.

The value of $p$ obtained from the best fit model is $1.4$, for a $q$ of $1.3$. The
jet break occurs at around half a day. We assumed $\nu_c$ to be
$\sim 2.5 \times 10^{13}$~Hz at the time of the break, below the optical bands,
to satisfy the observed $\alpha$ and $\delta$ in both x-ray and optical frequencies. The
synchrotron peak frequency $\nu_m$ is around $4 \times 10^{11}$~Hz at the time
of the jet break and the peak flux $f_{\nu_m}$ is $\sim 1.4$~mJy. The self absorption frequency $\nu_a$ cannot be constrained
using current observations. Our model requires additional extinction from the
host, with rest frame $A_v$ of $0.09$ corresponding to an $E_{B-V}$ of $0.04$
and a starburst type extinction law \citep{1997AJ....113..162C}. 

The derived total
energy of the burst is $3.6 \times 10^{49} n_0^{2/5}$~ergs, confined in an opening angle
of $2.3^{\circ} n_0^{1/10}$. The upper limit on $\xi$ is $10^4$. The polarisation lightcurve of this afterglow has been explained in terms of a structured jet \citep{2004A&A...422..121L}. The lightcurve from a structured jet viewed at an angle $\theta_0$ hardly differs from that of a homogeneous jet with half opening angle $\theta_0$ \citep{2002MNRAS.332..945R} (especially for a jet structure described by a $\theta^{-2}$ powerlaw). Hence we can still safely assume the shallow powerlaw model for the electron energy distribution within the jet, even though we are not using the structured jet calculations. However, The total energy calculations will be affected, if the energy distribution is not homogeneous within the jet. If we assume that our inferred value of $\theta_0$, which according to \citeauthor{2002MNRAS.332..945R} will be the viewing angle, is approximately equal to the half opening angle of the core of the structured-jet \citep{2002MNRAS.332..945R}, and if the actual extent of the jet is $90^{\circ}$, the energy inferred will be $\sim 9$ times smaller than the true energy (see \citeauthor{2002MNRAS.332..945R} for details).

The best fit model along with the observations are displayed in figure 6. The spectral parameters and physical parameters are listed in table 4 and table 5 respectively.
\subsection{GRB041006}
We have presented multiband modelling of this afterglow, which is yet another example of a $p<2$ electron distribution, in another paper \citep{2005BASI...33..487M}. We therefore do not describe this in detail here. We assume the cooling frequency ($\nu_c$) to be below the optical bands to satisfy  $\alpha$ of $0.5$ and $\delta$ in the range of $0.6 - 0.7$ simultaneously. There is no signature of steepening seen at the higher energy end of the spectrum from the available observations. Hence we place $\nu_i$ above the x-ray band. We compute the spectral evolution of the afterglow with these basic assumptions. For the sake of completeness, we list the spectral and physical parameters from our model in table 4 and table 5.

\section{Conclusions}

In GRB afterglows, as in other non-thermal sources, the shock
  accelerated electron spectrum at times assume a hard distribution \citep{1992ApJ...390..454H,1989MNRAS.239..401L}.
But almost all of the theoretical and
modelling work in GRB afterglow physics, by default, assume a single steep
power law for the distribution of electrons in the downstream plasma. The presence of a $p<2$ spectrum, in a minority of cases, has however not received a fair share of attention. Calculations to derive the physical parameters of the burst in such cases are often not done consistently. Early attempts to model GRB afterglows with hard electron energy spectrum had several loopholes. 

We have, in this paper, followed the approach of parametrising the temporal
evolution of $\gamma_i$ (thereby leaving room to account for different
possible physical processes that could determine $\gamma_i$) as $\gamma_i \propto \Gamma^q$ (B01) and obtaining the afterglow flux decay index for different values of $q$. We have obtained expressions to calculate the observables from the physical parameters of the system which in turn can be used to derive the latter. We present multiband modelling of three afterglows, assuming ultra-relativistic expansion, and estimated their physical parameters. 

For all these afterglows, we obtain good fits when $q \ge 1$. The inferred lower limit of $\xi$ is around $10^4$. Within the present understanding of particle acceleration physics, a mechanism which produces $q \ge 1$ and $\xi \sim 10^4$ is not known. However, future observations of GRB afterglows in the high energy range which can be achieved by upcoming satellites {\textit{GLAST}} and {\textit{ASTROSAT}} will shed more lights on these parameters.
For none of the three afterglows, the synchrotron self absorption frequency was well
constrained. This left us with four observables and five unknowns, so we
obtained the physical parameters as a function of the assumed value of ambient medium
density. 
Though all of these afterglows were bright in their $\gamma$-ray output with
isotropic equivalent energy in $\gamma$-rays $\sim 10^{52}$ -- $10^{53}$~erg, the total kinetic energy derived from multiband modelling is relatively low ($\sim 10^{49}$~erg). This is partly due to the narrow beaming angle derived from an early jet break (for all the jets, $\theta$ is roughly $2.5^{\circ}$). Perhaps kinetic energy being an order of magnitude less than the energy output in radiation could be a trait associated with the presence of hard electron energy spectrum. More afterglows and their detailed modelling is required to examine this possibility.
Another significant characteristic of all the three afterglows is a relatively
low value of the synchrotron cooling frequency. While for most afterglows
discussed in the literature, $\nu_c$ remain above optical bands longer than a
day after the burst, the three afterglows discussed here have, in our model, $\nu_c$ falling below the optical band within $3$~hours.

The origin of the hard electron distribution is not yet clear. Different physical processes such as diffusive shock acceleration \citep{2001MNRAS.328..393A}, cyclotron wave resonance \citep{1992ApJ...390..454H} etc. are beginning to be explored in detail in the context of relativistic shocks. Further developments in this area will hold the key to understanding the origin of the observed spectra of Gamma Ray Bursts and their afterglows.
%
%
%
%
\begin{table}
\caption{Temporal indices of the spectral parameters. For general $q$ and $s$}
\begin{tabular}{|c|c|c|} \hline
frequency & before jet break & after jet break \\ \hline
$\nu_m$ &$\frac{s+(s-6)p_1-2q(2-p_1)(s-3)}{2(4-s)(p_1-1)}$&$\frac{2q-p_1-qp_1}{p_1-1}$ \\
&& \\
$\nu_a (\nu_a < \nu_m < \nu_c) $&$\frac{s(10q-4-p_{1}-5p_{1}q)+15(-p_{1}+p_{1}q-2q+2)}{10(4-s)(p_{1}-1)}$&$-\frac{7p_1-5p_1q+10q-12}{10(p
_1-1)}$ \\
&& \\
$\nu_a (\nu_m < \nu_a < \nu_c) $&$\frac{s(2+p_{1}-4q+2p_{1}q)-6p_{1}-20+12q-6p_{1}q}{2(4-s)(p_{1}+4)}$&$\frac{(2q-4-p_1-qp_1)}{p_1+4} $\\
&& \\
$\nu_a (\nu_a < \nu_c < \nu_m) $ & $\frac{3}{5} + \frac{22}{5(s-4)}$   &  $-\frac{2}{3}$  \\
$\nu_a (\nu_c < \nu_a < \nu_m) $ & $\frac{6-s}{3(s-4)}$   &  $-\frac{6}{5}$  \\
$\nu_i$&$\frac{s(1+2q)-6(q+1)}{2(4-s)}$&{\scriptsize{$-(1+q)$}} \\
&& \\
$\nu_c$&$\frac{3s-4}{2(4-s)}$&{\scriptsize{$0$}}\\
&& \\
$f_{\nu_{m}}$&$-\frac{s}{2(4-s)}$&{\scriptsize{$-1$}} \\ \hline
\end{tabular}
\end{table}
\begin{table}
\caption{The spectral indices ($\delta$) and lightcurve decay indices
  ($\alpha_1$; before jet break, $\alpha_2$; after jet break) for various
  spectral regimes in slow cooling phase. Note that $\alpha$ depends upon the
  value $q$ assumes. The expressions assume forms similar to those in $p > 2$
  case, if $q$ is set to unity. 
}
\begin{tabular}{|r|c|c|l|}\hline
spectral segment &$\delta$ & $\alpha_{1}$ (ISM,WIND)&$\alpha_{2}$\\ \hline
$\nu <\nu_{a} < \nu_{m} <\nu_{c}$ &  &  &   \\ 
& $2$ & $-\frac{(10-7p_{1}+3p_{1}q -6q)}{8(p_{1}-1)}$,   $\frac{6-5p_{1}+p_{1}q-2q}{4(1-p_{1})}$  &$\frac{3p_{1}-6-3p_{1}q+6q}{6(p_{1}-1)} $  \\ 
$\nu <\nu_{m} < \nu_{a} <\nu_{c}$ &  &  &   \\ \hline

$\nu_{a} <\nu < \nu_{m} <\nu_{c}$ &  &  &   \\
 &$\frac{1}{3}$ &$\frac{p_{1}+p_{1}q-2q}{4(P_{1}-1)} , \frac{2-p_{1}+p_{1}q-2q}{6(P_{1}-1)}$ &$\frac{-2p_{1}+3-2q+qp_{1}}{3(p_{1}-1)}$ \\ \hline

&&& \\
$\nu_{m} <\nu < \nu_{a} $ &$\frac{5}{2}$  &$\frac{5}{4}, \frac{7}{4}$  & $1$  \\ 
&&& \\ \hline

$\nu_{m} <\nu < \nu_{i} <\nu_{c}$ &  &  &   \\ 
&$-\frac{(p_{1}-1)}{2}$ & $-\frac{3}{8}(p_{1}+p_{1}q-2q),
\frac{1}{4}(2q-p_{1}q-2p_{1}-1)$ & $-\frac{2(q-1)-p_{1}(1+q)}{2}$ \\
$\nu_{m} <\nu < \nu_{c} <\nu_{i}$ &  &  &   \\ \hline

&&& \\
$\nu_{m} <\nu_{i} < \nu <\nu_{c}$ &$-\frac{(p_{2}-1)}{2}$
&$-\frac{3}{8}(p_{2}+p_2q-2q) , \frac{1}{4}(2q-p_{2}q-2p_2-1) $&$-\frac{2(q-1)-p_{2}(1+q)}{2}$   \\  
&&& \\ \hline

&&& \\
$\nu_{m} <\nu_{c} < \nu <\nu_{i}$ &$-\frac{p_{1}}{2}$
&$\frac{1}{8}(6q-3p_1-3p_1q-2),\frac{1}{4}(2q-p_{1}q-2p_{1})$ & $-\frac{2(q-1)-p_{1}(q+1)}{2}$   \\   
&&& \\ \hline

$\nu_m < \nu_{i} <\nu_{c} < \nu $ &  &  &   \\ 
 & $-\frac{p_{2}}{2}$&$\frac{1}{8}(6q-3p_2-3p_2q-2), \frac{1}{4}(2q-2p_{2}-p_{2}q) $ &$-\frac{2(q-1)-p_{2}(q+1)}{2}$ \\
$\nu_m < \nu_{c} <\nu_{i} < \nu $ &  &  &   \\ \hline
\end{tabular}
\end{table}
\begin{table}
\caption{Same as table 2, but for fast cooling phase. After $\nu$ goes above both $\nu_c$ and $\nu_m$, the respective positioning of these frequencies does not affect lightcurve slope and the indices will be the same as that of the corresponding slow cooling regime.}
\begin{tabular}{|r|c|c|l|}\hline
spectral segment &$\delta$ & $\alpha_{1}$ (ISM,WIND)&$\alpha_{2}$\\ \hline
$\nu<\nu_a<\nu_c)$ & 2 & 1 , 2 & 1/9\\
$\nu_a< \nu <\nu_c)$ & 1/3 & 1/6 , -2/3 & -1\\
$\nu<\nu_c<\nu_a)$ & 2 & 1 , 2 & 13/5\\
$\nu_c< \nu <\nu_a)$ & 5/2 & 5/4 , 7/4 & 13/5 \\
$(\nu_a,\nu_c)<\nu<\nu_m$ & -1/2 & -1/4 , -1/4 & -1\\ \hline
\end{tabular}
\end{table}
\begin{table}
\caption{Fit parameters of the three modelled afterglows, given around the
  time of jet break. 
}
\begin{tabular}{lccc} \hline
Fit Parameters& GRB010222& GRB020813& GRB041006 \\ \hline
$p_1$ & $1.47_{-0.003}^{+0.004}$ & $1.40^{+0.007}_{-0.004}$ & $1.29- 1.32$ \\ 
$p_2$ & $ 2.04_{+1.76}^{-0.01}$ & $ \sim 2.1$ & $> 2.2$ \\
$q$& $1.3 \pm 0.06$& $1.3 \pm 0.05$ & $0.95 - 1.14$ \\
$\nu_m$~Hz & $ 2.24^{+9.4}_{-0.65} \times 10^{11}$ & $3.99^{+1.58}_{-0.95} \times 10^{12}$ & $ (1.2 - 3.0) \times 10^{12}$ \\
$\nu_c$~Hz & $ 9.03^{+0.37}_{0.36} \times 10^{13}$ & $ 2.33^{+0.14}_{-0.28} \times 10^{13}$  &  $ (1.0 - 2.0) \times 10^{14}$ \\
$\nu_i$~Hz & $ > 10^{19}$ & $ > 5 \times 10^{19}$ &  $ > 2.4 \times 10^{20}$ \\
$f_p$~mJy & $1.037^{+0.01}_{-0.108}$ & $1.35^{+0.025}_{-0.065}$ & $(0.37 - 0.49)$ \\
$t_j$~day & $0.56^{+0.035}_{-0.033}$ & $0.48 \pm 0.03$  &$0.17 - 0.24$ \\
$E_{(B-V)}$~(host) & $0.035^{+0.005}_{-0.0035}$~mag & $0.03^{+0.006}_{-0.003} $~mag &  $0.01 - 0.05$~mag \\
&&& \\
Host Gal. B band & $25.64^{+0.5}_{-0.25}$~mag & -- & -- \\
" V band & $26.29^{+0.25}_{-0.5}$~mag & -- &  --\\
" R band & $25.83^{+0.25}_{-0.3}$~mag & -- & --\\
" I band & $25.59 \pm 0.25$~mag & -- & --\\
" $8.46$~GHz & $25^{+25}_{-19}\mu$Jy& -- &  --\\
" $4.86$~GHz & $20^{+59}_{-10}\mu$Jy& -- & -- \\
\end{tabular}
\end{table}
\begin{table}
\caption{Derived physical parameters for the three afterglows. Since $\nu_a$
  was not well constrained in all the cases, the parameters are presented as a
  function of the ambient density $n_0$, normalised to $1$~atom/cc.}
\begin{tabular}{cccc} \hline
physical parameters & GRB010222 & GRB020813 & GRB041006 \\ \hline
$\epsilon_e \, n_0^{-\frac{p_1}{20}}$ &  $\sim 1.0$ &  $\sim 1.0 $ & $\sim 0.8 $  \\
&&&\\
$\epsilon_B \, n_0^{\frac{3}{5}}$ & $0.027^{+0.001}_{-0.002}$ & $0.1^{+0.004}_{-0.007}$   & $0.07 - 0.14$  \\
&&&\\
$\xi \, n_0^{-\frac{1}{20}}$  & $12.0^{+11.5}_{-3.9} \times 10^{4}$ & $ > 5.7 \times 10^{4}$ & $ > 2.0 \times 10^{4}$ \\
&&&\\
$E_{\rm{iso}} \, n_0^{-\frac{1}{5}}$~ergs &  $5.83^{+0.14}_{-1.0} \times 10^{52}$ & $3.22^{+0.076}_{-0.175} \times 10^{52}$ & $(2.0-4.0) \times 10^{51}$ \\
&&&\\
$\theta_j \, n_0^{-\frac{1}{10}}$~deg. & ${2.0^{\circ}} \pm 0.008$ & ${2.3^{\circ}} \pm 0.05$ & $1.7^{\circ} - 2.8^{\circ}$ \\
&&&\\
$E_{\rm{tot}} \, n_0^{-\frac{2}{5}}$~ergs & $3.60 \pm 0.002 \times 10^{49}$ & $2.2 ^{+0.4}_{-1.5}\times 10^{49}$ &  $(1.4 -3.4) \times 10^{48}$ \\ \hline
\end{tabular}
\end{table}
%
%
%
%
%
\begin{figure}           
\centering
\subfigure{%
\put(15,100){\rotatebox{90}{log($\nu_a$/Hz)}}
\put(78,0){log($(t-t_0)$/day)}
\includegraphics[scale=0.6]{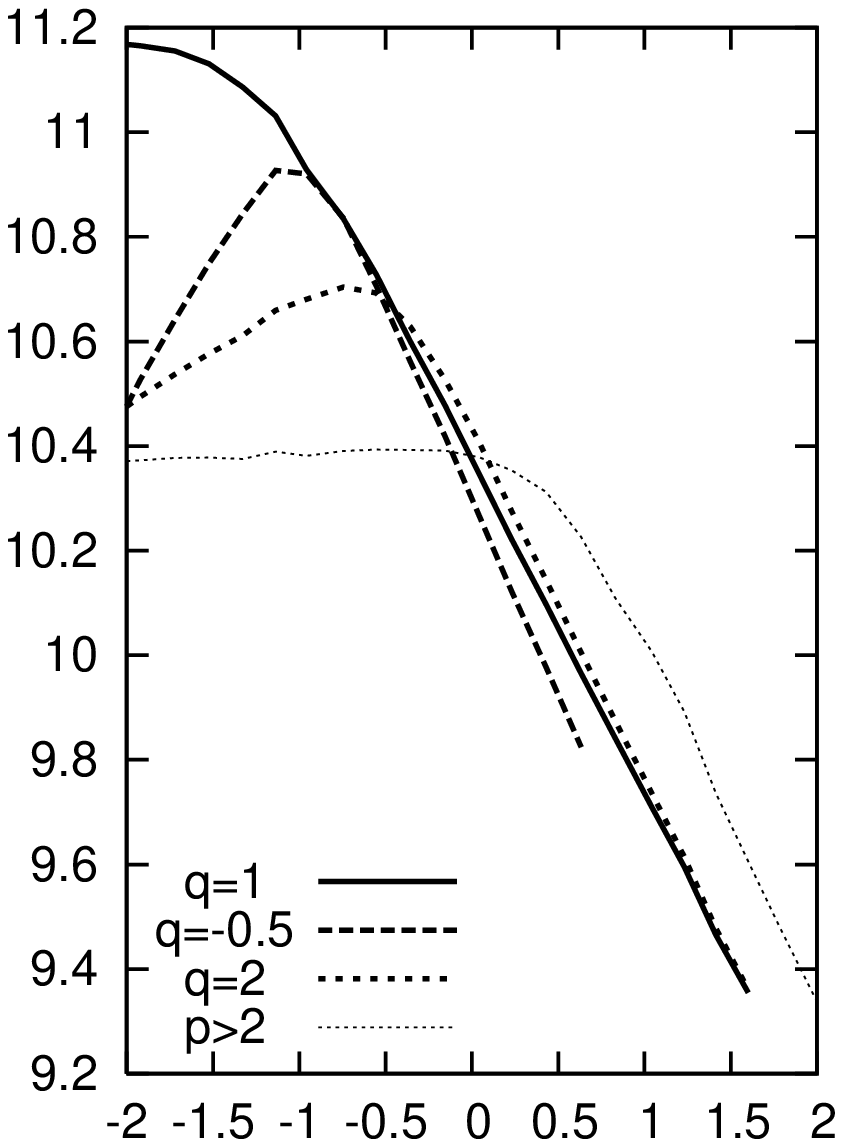}} \qquad
\subfigure{%
\put(15,100){\rotatebox{90}{log($\nu_m$/Hz)}}
\put(78,0){log($(t-t_0)$/day)}
\includegraphics[scale=0.6]{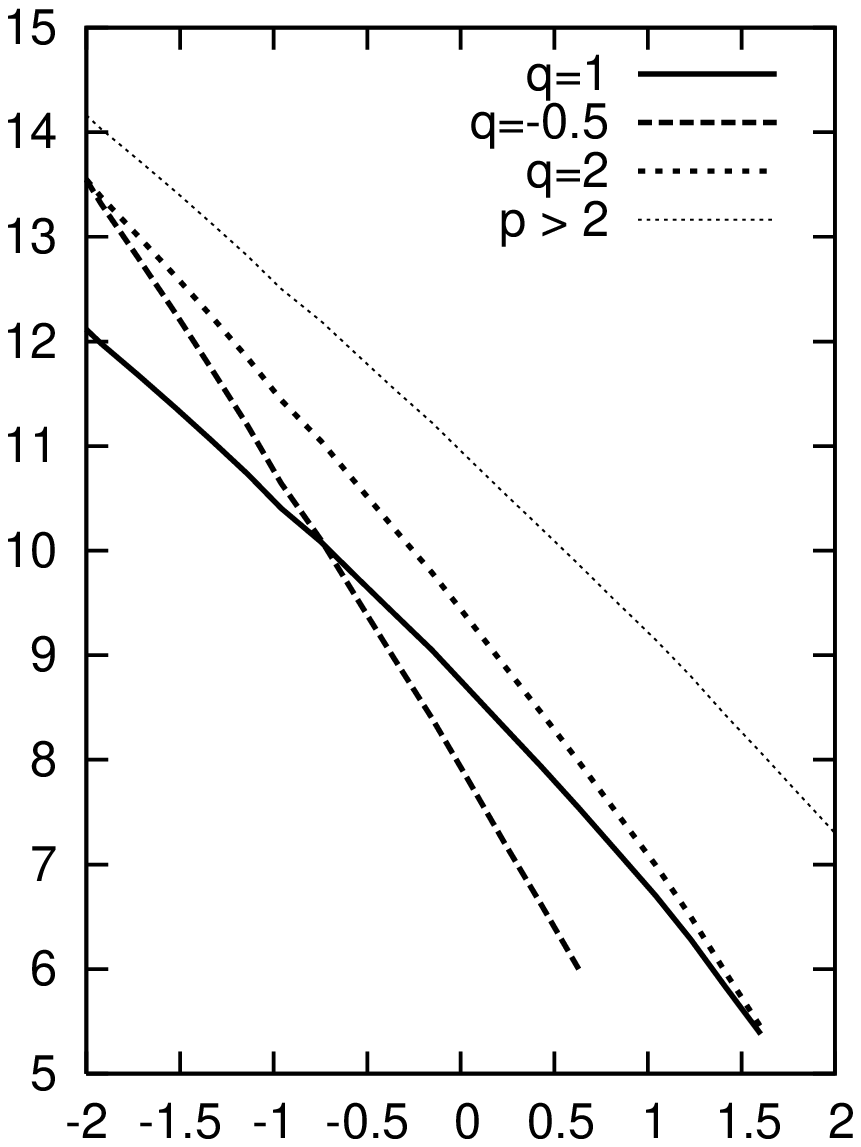}} \\
\subfigure{%
\put(15,100){\rotatebox{90}{log($\nu_i$/Hz)}}
\put(78,0){log($(t-t_0)$/day)}
\includegraphics[scale=0.6]{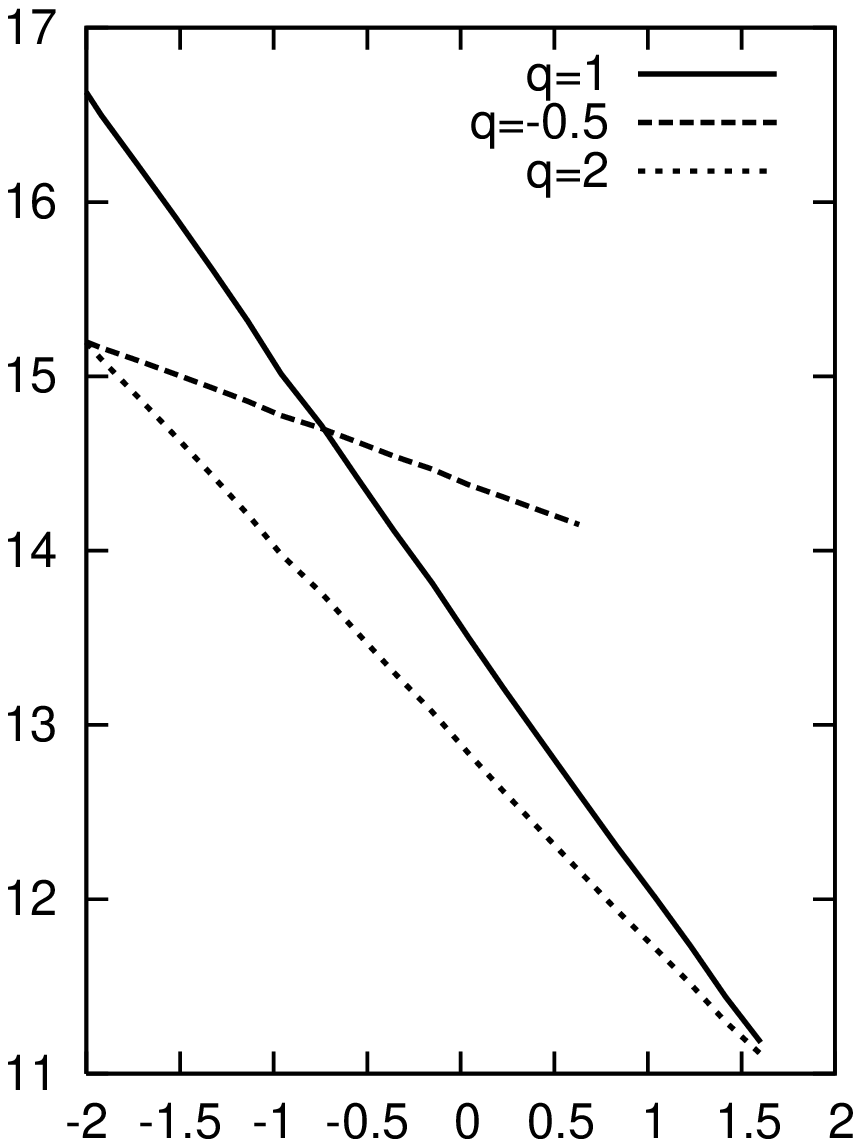}}
\caption{Evolution of spectral breaks $\nu_a$ (top left), $\nu_m$ (top right)
  and $\nu_i$ (bottom) for different values of $q$. For comparison, result of
  a single power law with $p =2.2$ is also shown (thin line). $\nu_i$ is not
  relevant for the $p>2$ case however. The parameters used in calculating the
  curves are: $z = 1$, a spherical outflow of isotropic equivalent energy
  $10^{51}$ ergs and initial lorentz factor $350$ in a homogeneous ambient
  medium of density 0.1 atom/cc. The shock microphysics parameters are:
  $\epsilon_e = 0.1$, $\epsilon_B = 0.01$, $p_1=1.5$, $p_2=2.2$ and $\xi =
  2000$.
}

\end{figure}
\begin{figure}
\centering
\subfigure{
\put(15,100){\rotatebox{90}{log($f_\nu$/mJy)}}
\put(78,0){log($(t-t_0)$/day)}
\includegraphics[scale=0.6]{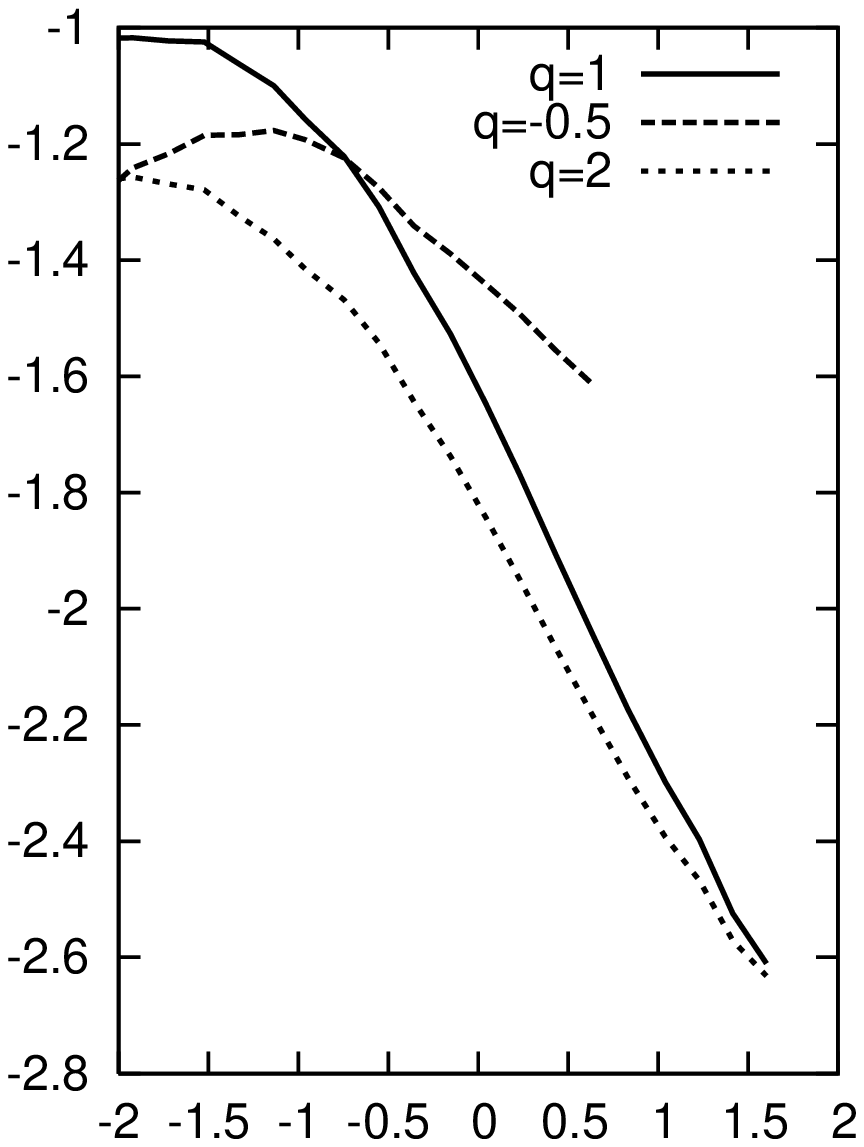}} \qquad
\subfigure{
\put(15,100){\rotatebox{90}{log($f_\nu$/mJy)}}
\put(78,0){log($(t-t_0)$/day)}
\includegraphics[scale=0.6]{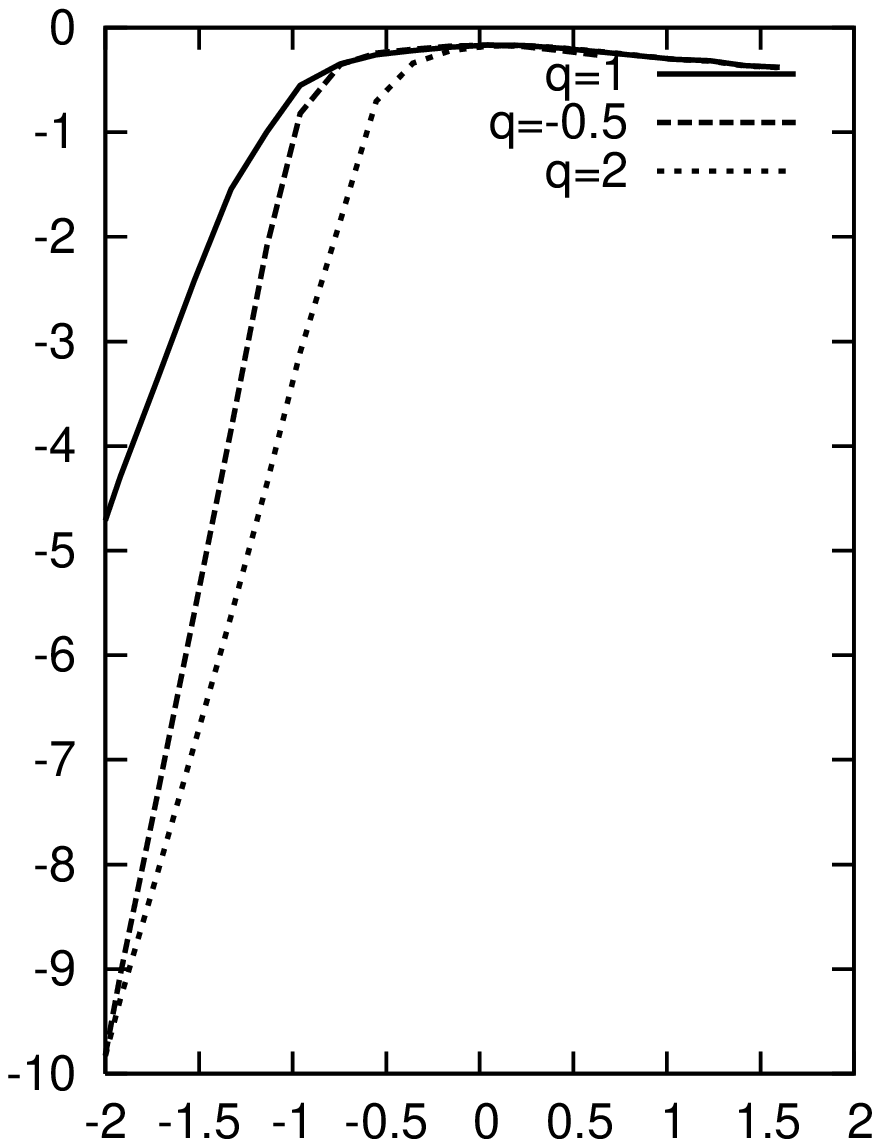}} \\
\subfigure{
\put(15,100){\rotatebox{90}{log($f_\nu$/mJy)}}
\put(78,0){log($(t-t_0)$/day)}
\includegraphics[scale=0.6]{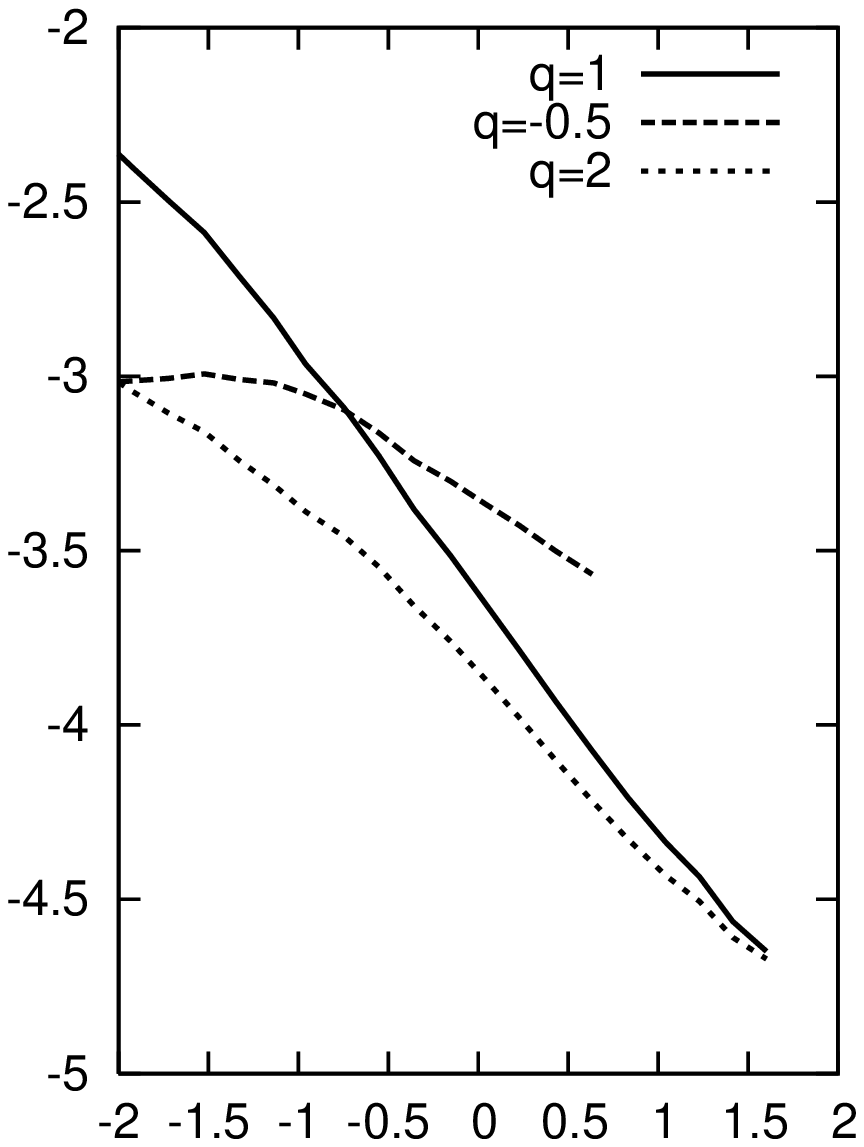}}
\caption{(left top) Sample model optical lightcurve ($4 \times 10^{14}$~Hz), (right top) x-ray lightcurve ($10^{18}~Hz$) and (bottom) radio lightcurve for $22$~GHz  for three different values of $q$.}
\label{fig-c2-04}
\end{figure}
\begin{figure}
\includegraphics[scale=0.6,angle=-90.]{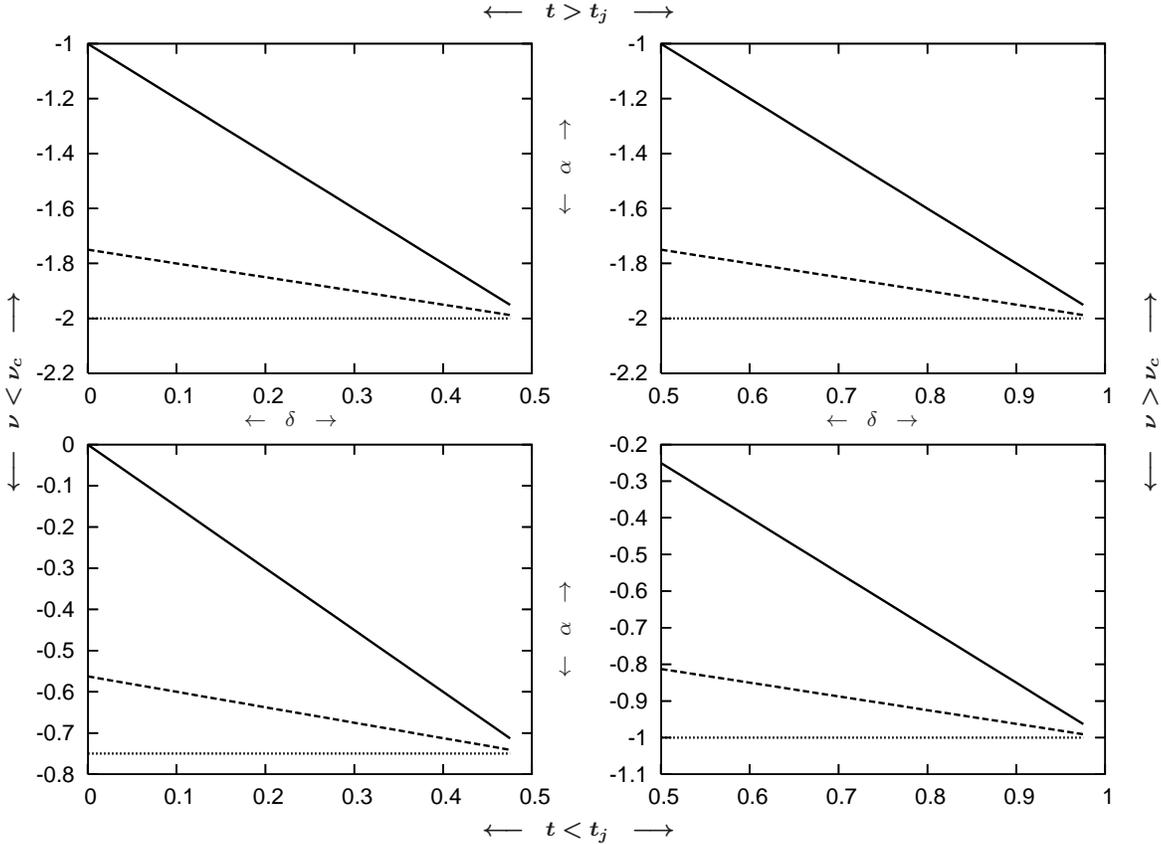}
\put(-430,-180){\rotatebox{90}{\boldmath $\longleftarrow \; \; \nu < \nu_c \; \; \longrightarrow$}}
\put(0.,-180){\rotatebox{90}{\boldmath $ \longleftarrow \; \; \nu > \nu_c \; \; \longrightarrow $}}
\put(-250,-310){\boldmath $ \longleftarrow \; \; t < t_j \; \; \longrightarrow $}
\put(-250,0.){\boldmath $ \longleftarrow \; \; t > t_j \; \; \longrightarrow $}
\put(-340.,-155){$\leftarrow \; \; \delta \; \; \rightarrow$}
\put(-120.,-155){$\leftarrow \; \; \delta \; \; \rightarrow$}
\put(-220,-250){\rotatebox{90}{$\leftarrow \; \; \alpha \; \; \rightarrow$}}
\put(-220,-75){\rotatebox{90}{$\leftarrow \; \; \alpha \; \; \rightarrow$}}
\caption{The $\alpha$ -- $\delta$ closure relations for various values of $q$. The left panel shows the closure relations when the observing frequency is below the $\nu_c$, the right panel is for $\nu > \nu_c$. In the bottom panels $\alpha$ is calculated before jet break. In the top panel, post jet break $\alpha$ values are presented. Solid line is for $q = 1$, dotted line is for $q=-0.5$ and dashed line is for $q=-1$. For $q=1$, the standard $p>2$ scaling is recovered. Note that for $q=-0.5$, the minimum possible value of $\alpha$ is $1.75$. For $q=-1$, $\alpha$ does not depend on $\delta$.}
\end{figure}
\begin{figure}
\begin{center}
\leavevmode
\includegraphics[width=8.cm,height=10.cm,angle=270]{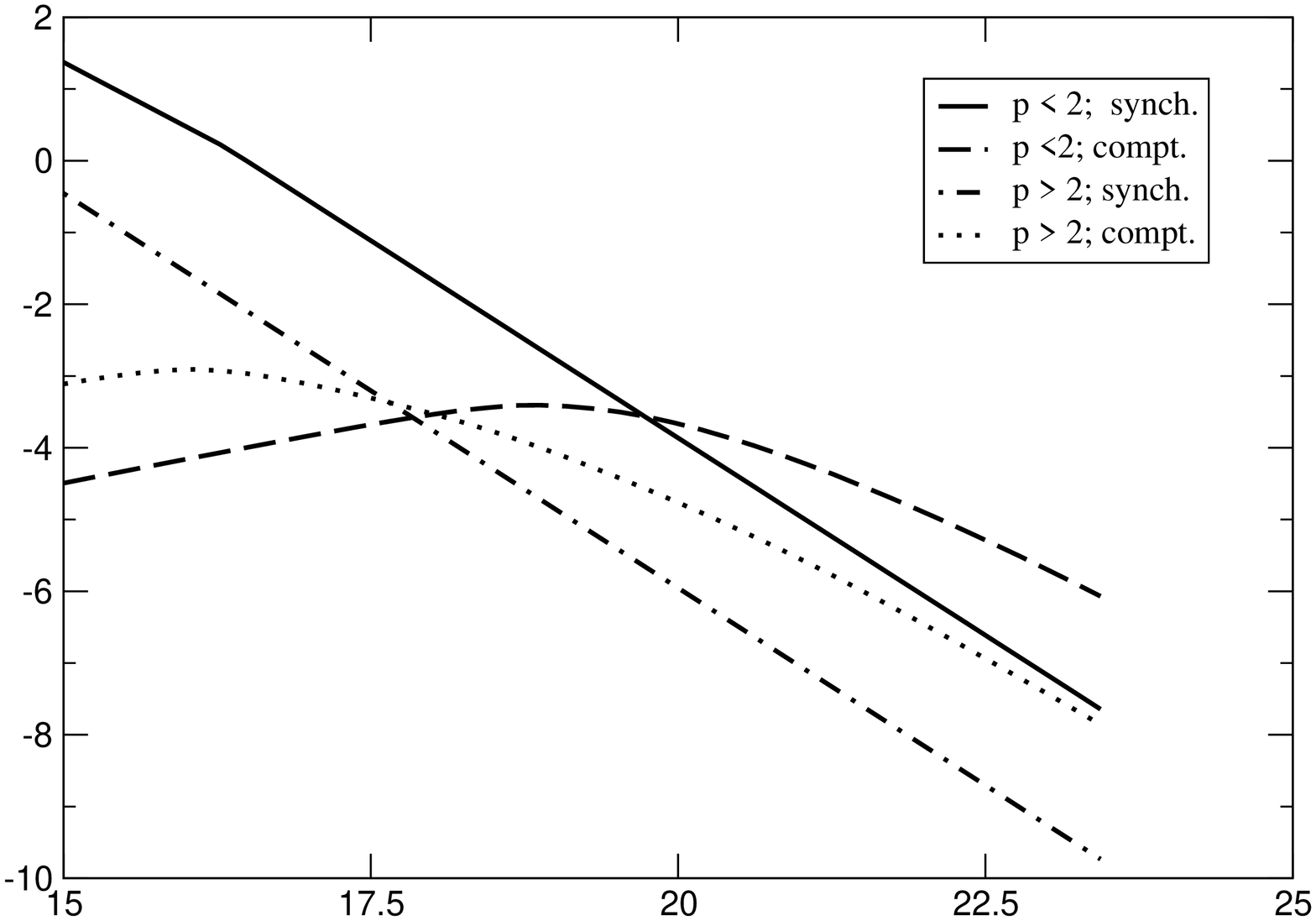}
\put(-280,-150){\rotatebox{90}{log $f_\nu$/mJy}} 
\put(-160,-220){log $\nu$/Hz} 
\caption{The predicted compton contribution from hard electron energy spectrum, in comparison with that from a steep spectrum. For frequencies less than $10^{19}$~Hz, the contribution from SSA is rather low for $p < 2$ spectrum. The parameters used for calculation are, $\enorm = 10^2$, $n = 100$, $\epsilon_e = 0.3$ and $\epsilon_B = 10^{-3}$. For 
hard spectrum $p_1 = 1.8$, $p_2 = 2.2$, $q=1$ and $\xi = 5000$ are used, and
for steep spectrum a $p$ of $2.2$ is used. The displayed spectra are for $\sim
5$~days post-burst.}
\end{center}
\end{figure}
\begin{figure}
\centering
\subfigure[]{\includegraphics [width=8cm]{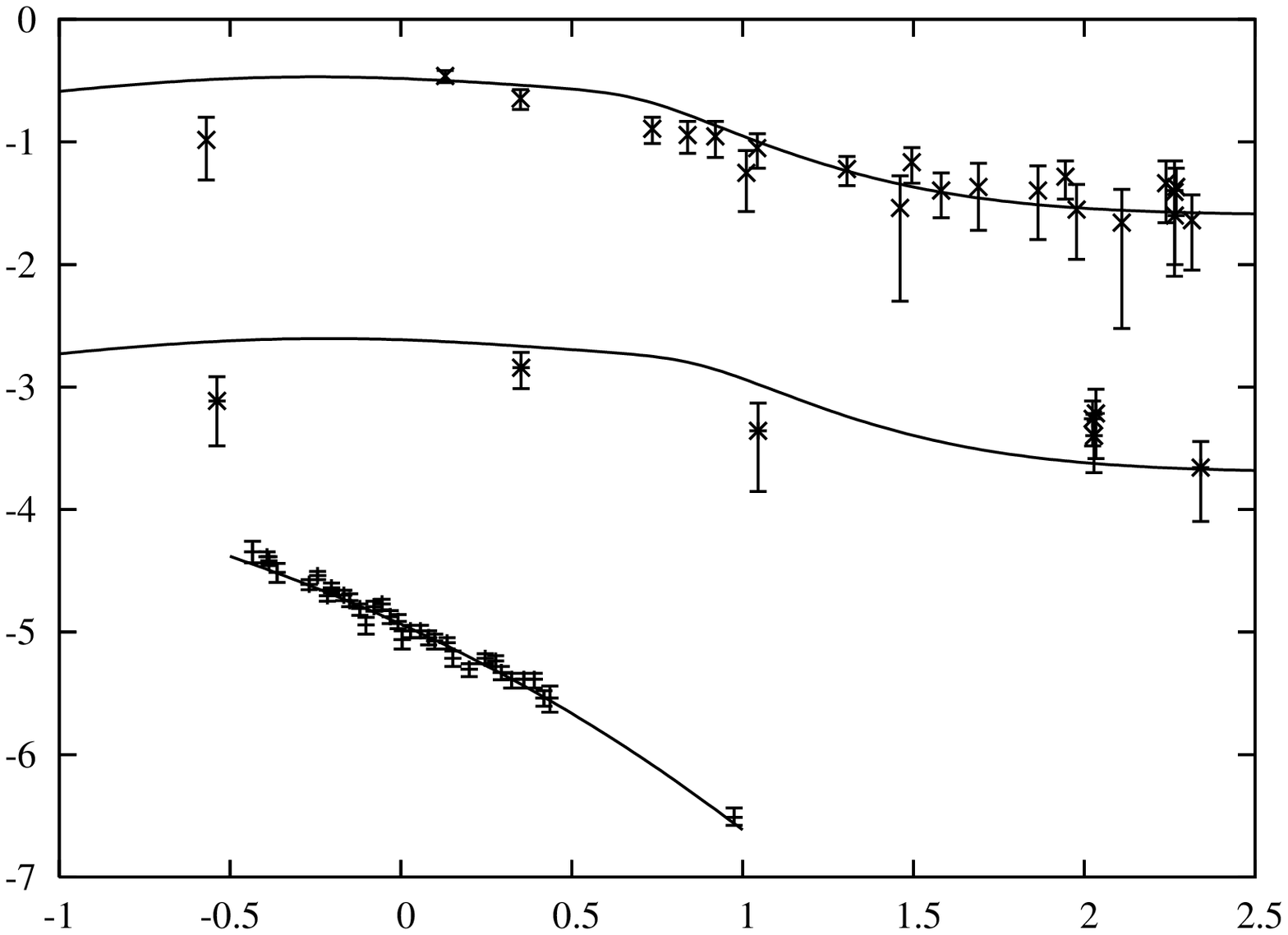}} \qquad
\put(-153,-2){log($(t-t_0)$/day)}
\put(-231,57){\rotatebox{90}{log $f_{\nu}$/mJy}}
\put(-59,137){\footnotesize{8GHz}}
\put(-150,107){\footnotesize{4GHz/100}}
\put(-120,47){\footnotesize{Xray/10}}
\subfigure[]{\includegraphics [width=8cm]{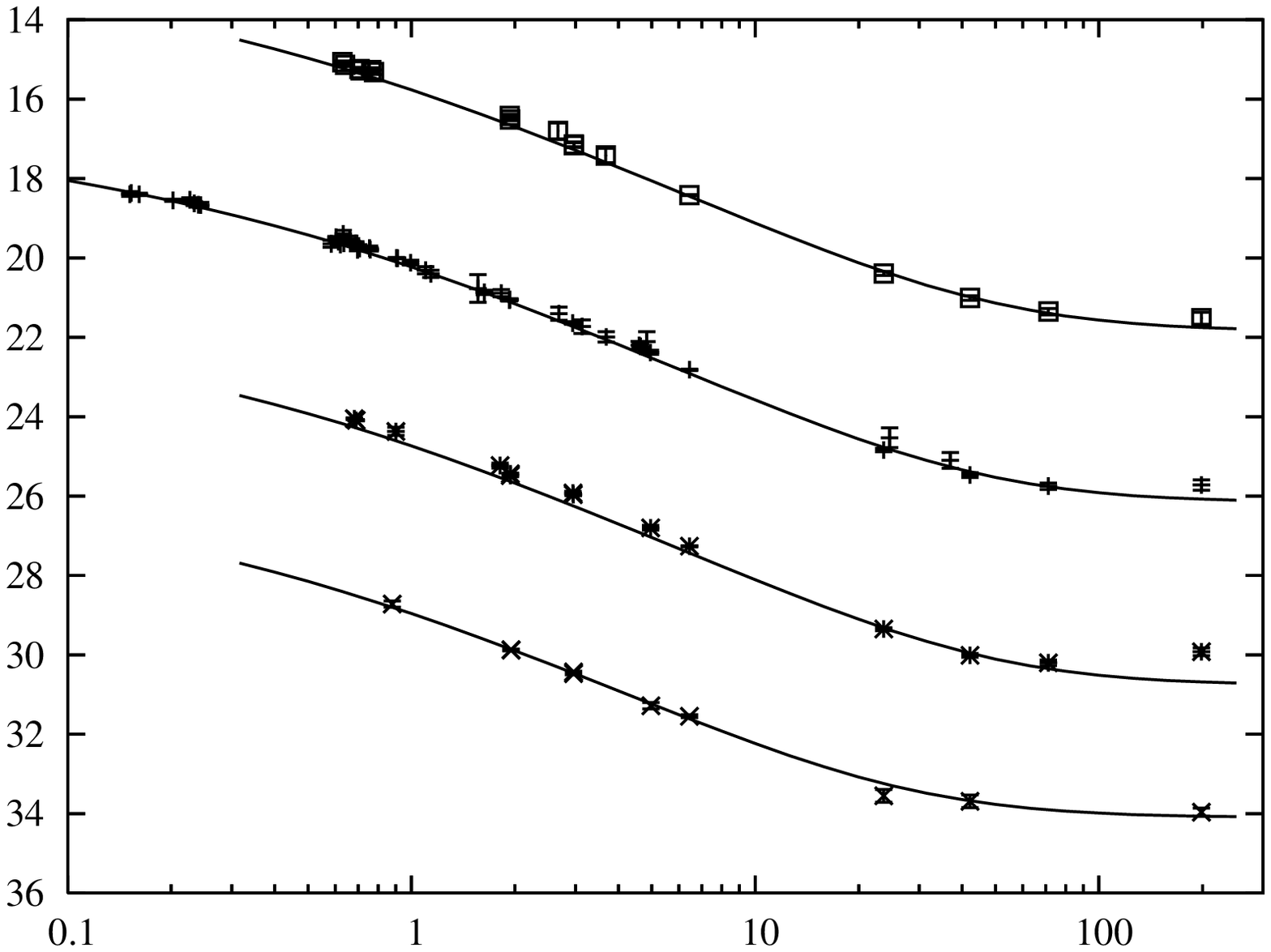}} 
\put(-153,-2){log ($(t-t_0)$/day)}
\put(-227,57){\rotatebox{90}{Magnitude}}
\put(-94,127){\footnotesize{Iband-4}}
\put(-99,100){\footnotesize{Rband}}
\put(-90,70){\footnotesize{Vband+4}}
\put(-99,48){\footnotesize{Bband+8}}
\\
\subfigure[]{\includegraphics [width=8cm]{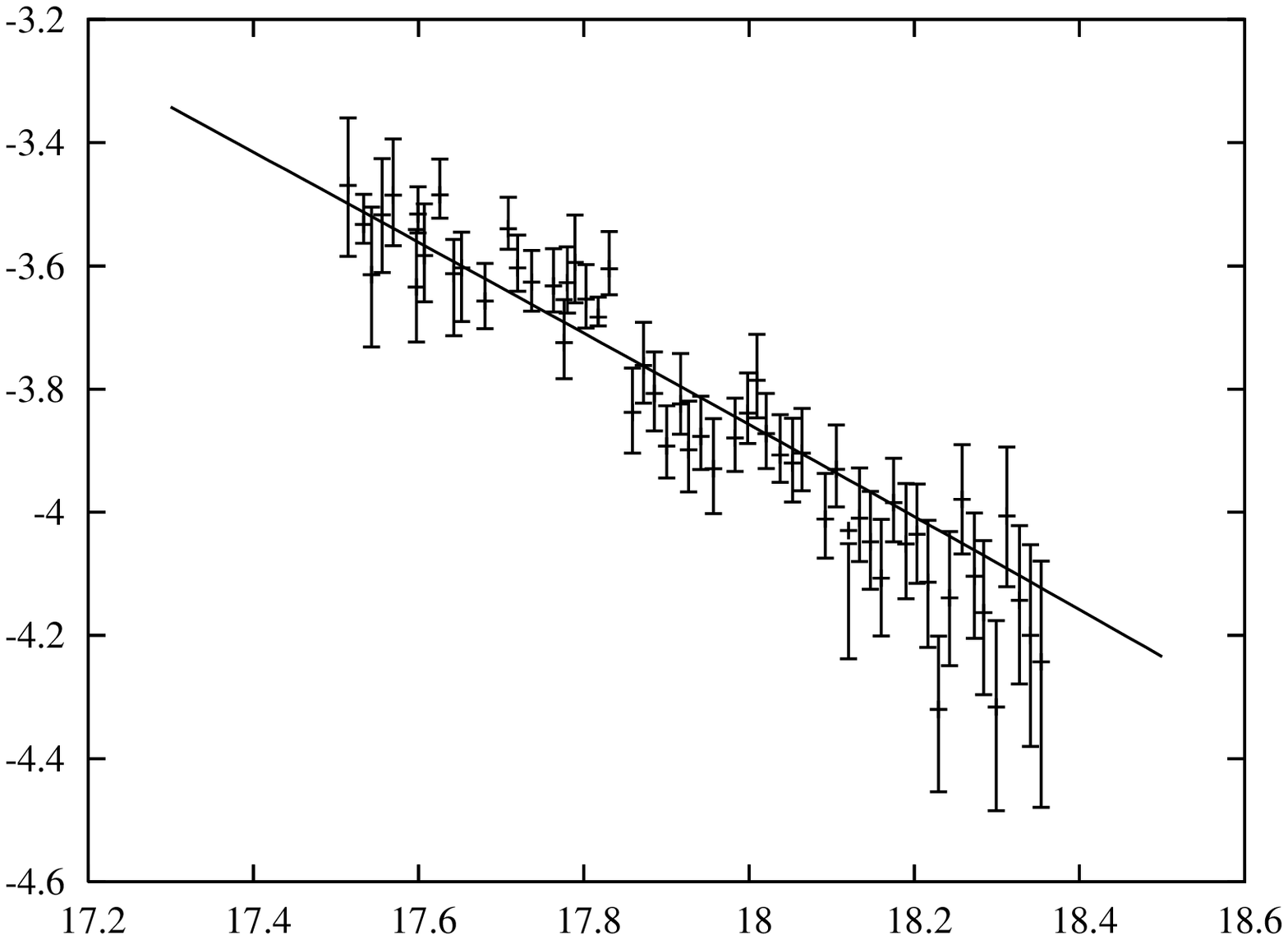}}
\put(-123,-2){log $\nu$/Hz}
\put(-231,57){\rotatebox{90}{log $f_{\nu}$/mJy}}
\put(-70,100){\footnotesize{at $\sim$ 1 day}}
\caption{Multiband model fits for GRB010222. Points : observed data. Solid line : our model. (a) Radio and x-ray lightcurves. The $4$~GHz lightcurve and the $10^{18}$~Hz x-ray lightcurve are offset by $0.01$ and $0.1$~mJy respectively for the ease of viewing. The flattening seen in radio lightcurves (panel a) are due to the flux of the starburst host SMMJ14522+4301 (see text for details). (b) Optical BVRI lightcurves, appropriately offset to avoid clustering. (c) X-ray spectrum at $\sim 1$~day from BeppoSAX along with the model.}
\end{figure}
\begin{figure*}
\label{fig-020813}
\includegraphics[width=8cm]{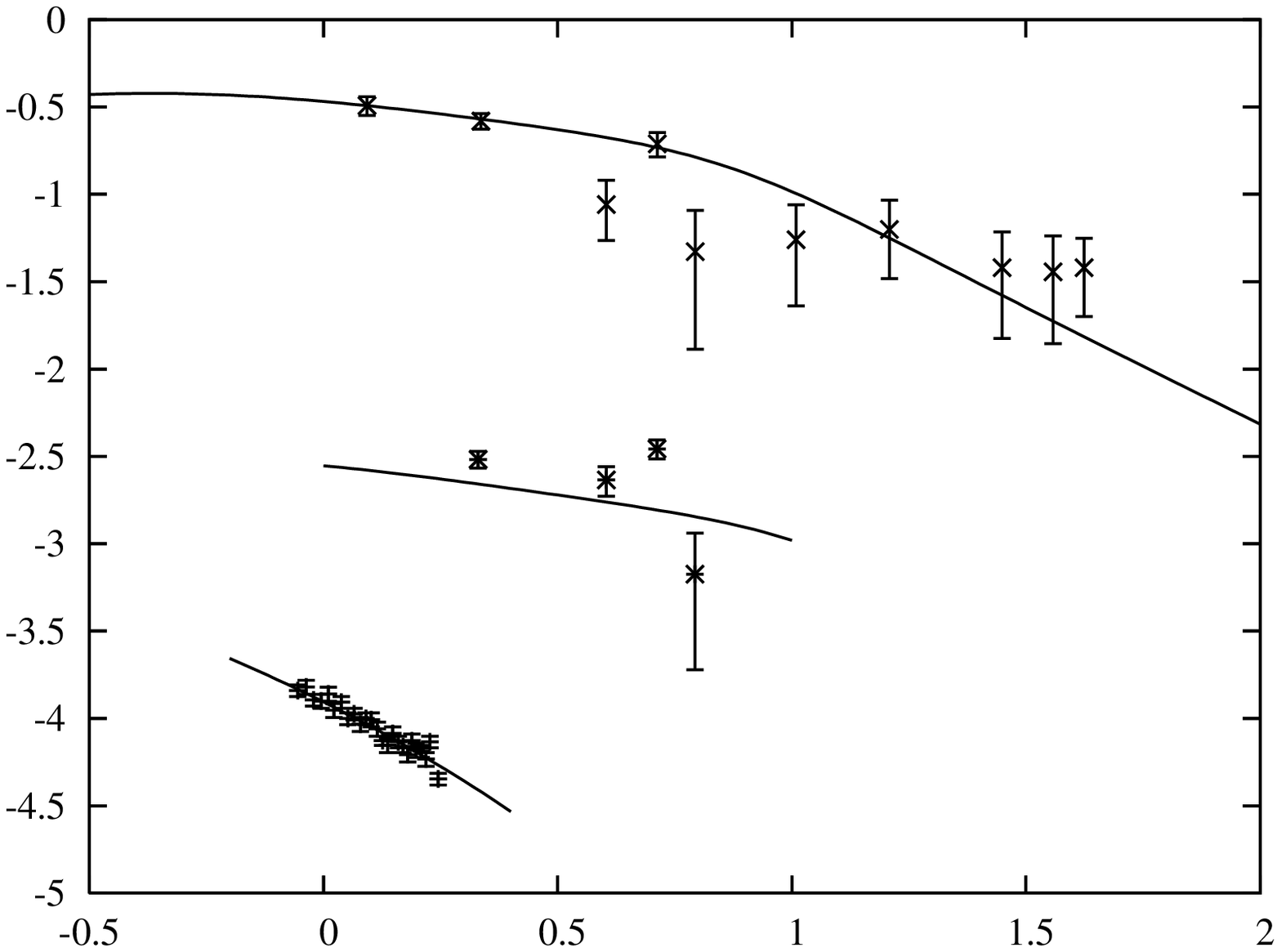}
\put(-228,57){\rotatebox{90}{log $f_{\nu}$/mJy}}
\put(-153,-2){log ($(t-t_0)$/day)}
\put(-52,137){\footnotesize{8GHz}}
\put(-165,95){\footnotesize{4GHz/100}}
\put(-140,47){\footnotesize{Xray/10}}
\includegraphics[width=8cm]{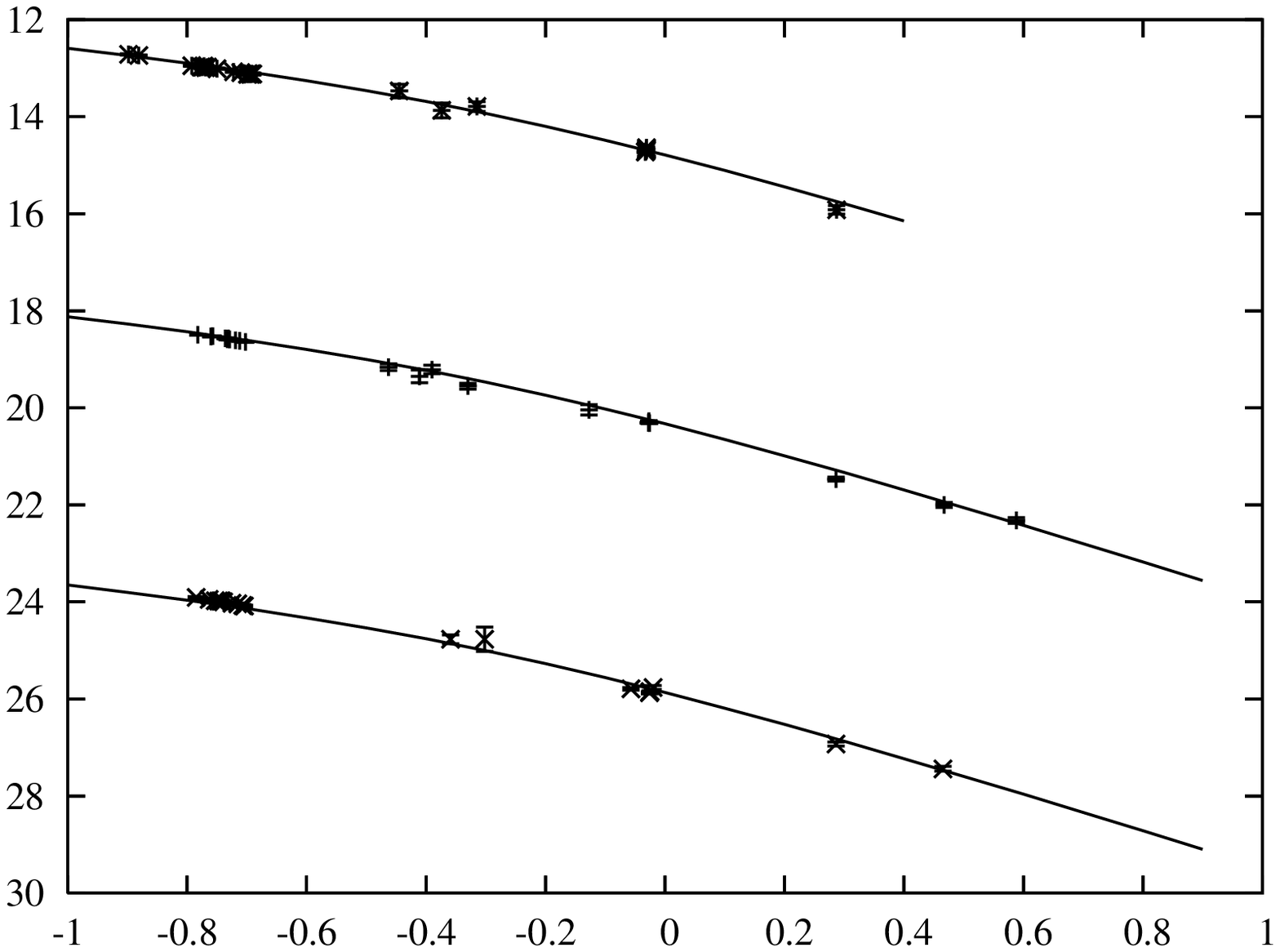}
\put(-227,57){\rotatebox{90}{Magnitude}}
\put(-153,-2){log($(t-t_0)$/day)}
\put(-94,137){\scriptsize{Iband-5}}
\put(-90,100){\scriptsize{Rband}}
\put(-99,52){\scriptsize{Vband+5}}
\caption{GRB020813: Best fit model along with the observations. (i) The top two
  curves in the left side panel are radio flux in $8.46$~GHz and $4.86$~GHz
  respectively. For ease of viewing, $4.86$~GHz flux is multiplied by
  $0.01$~mJy. The late time flattening in the $8$~GHz data is not due to the
  presence of any host. Such flattening is seen in the radio afterglows beyond
  a few days past the burst, and is suspected to be some non-standard
  behaviour (see Frail et al. 2004) which is not taken care of by our
  code. The bottom curve in this panel is the x-ray lightcurve at $1.2 \times
  10^{18}$~Hz. (ii) The right panel displays multiband optical lightcurves. I band is offset by $-5$ magnitudes while V band is off set by $+5$ magnitudes.}
\end{figure*}
\newpage
\clearpage
\appendix
\section{Calculation of the lateral velocity of the jet}
The adiabatic sound velocity is defined as, $c_{s} = dP / d\rho$ where $P$ is
the gas pressure and $\rho$ is the mass density. \citet{1939isss.book.....C} derives the thermal energy density $U_{k}$ of a mono-atomic gas to be,
\be
U_{k}= n\left[\frac{3K_{3}(\Theta) + K_{1}(\Theta)}{4K_{2}(\Theta)} -1\right]m_{1}c^{2},
\ee
where $n$ is the particle number density in the gas and $m_{1}$ is mass of
a single particle.
$\Theta = m_{1}c^{2}/k_{B}T$, where $T$ is the temperature of the gas.
$K_{n}(\Theta)$ is the modified Bessel function of order $n$.
In terms of temperature, thermal energy density is usually expressed as,
$n\alpha(T)k_{B}T$, where $\alpha(T)$ parametrises the temperature
dependence. It follows from the two expressions that,
\be
\alpha(T) =\Theta\left[\frac{3K_{3}(\Theta) +K_{1}(\Theta)}{4K_{2}(\Theta)}-1\right]
\ee
In the non-relativistic regime, $\alpha(T)$ approaches the familiar value $3/2$ and in the relativistic limit, it tends to $3$.
For a blast wave downstream plasma, with single particle rest mass $m_1$, the average thermal energy per particle
$\alpha(T)k_{B}T$ can be written as $(\Gamma - 1)m_{1}c^{2}$.
i.e.,
\be
m_{1}c^2\left[\frac{3K_{3}(\Theta) +K_{1}(\Theta)}{4K_{2}(\Theta)}-1\right] = (\Gamma - 1)m_{1}c^{2}
\ee
from which we identify $(3K_{3}(\Theta) +K_{1}(\Theta)) /4K_{2}(\Theta)$ with $\Gamma$.
Temperature of the gas can be solved for, in terms of $\Gamma$ by inverting this relation.

But the total energy density is independent of the dynamic regime of the gas and is given by, $u = \rho c^{2} = (U_{k}+nm_{1}c^{2})/V$
where $\rho$ is the total (rest+inertial) mass density.
Using this expression we obtain,
\be
\frac{\rho}{P} = \Theta\frac{3K_{3}(\Theta) + K_{1}(\Theta)}{4K_{2}(\Theta)} = \Theta\Gamma
\ee
which gives the sound velocity in the downstream in terms of $\Gamma$ as,
\be
\left[ \frac{c_{s}}{c}\right]^{2} = \frac{1}{\Theta\Gamma} \label{eq-c2-08}
\ee
Let us examine the limiting values of the above expression and check the consistency.
In the non-relativistic limit, $k_{B}T \ll m_1c^2$ ie., $\Theta \gg 1$, the Bessel function takes the form
\be
K_{n}(\Theta) = \left[\pi \over {2\Theta}\right]^{1 \over 2}\exp{(-\Theta)}\left[1+\frac{4n^2 -1}{8\Theta}\right]  \label{eq-c2-10}
\ee
Substituting eqn. (\ref{eq-c2-10}) in eqn. (\ref{eq-c2-08});
\be
c_{s}^{2}=\frac{k_{B}T}{m_{1}}\frac{4\left[1+ \frac{15}{8\Theta}\right]}{3\left[1+\frac{35}{8\Theta}\right]+\left[1+\frac{3}{8\Theta}\right]}
\ee
Neglecting terms of the order of $1 / \Theta$, expression for sound velocity in a non-relativistic gas is reduced to
\be
c_{s}^{2}=\frac{k_{B}T}{m_{1}} \label{eq-c2-11}
\ee
Now, in the relativistic limit, ie., when $\Theta \ll 1$
The limiting expression for Bessel function is,
\be
K_{n}(\Theta) =\frac{1}{2}\frac{(n-1)!}{(\frac{\Theta}{n})^{n}}
\ee
Substituting the above expression in (12), and neglecting terms $O(\Theta^2)$, we get for the sound velocity in a relativistic gas,
\be
c_{s}^{2}=\frac{k_{B}T}{m_{1}}\frac{8\Theta}{24} = {c^2 \over 3}
\ee
We calculate the the lateral velocity of matter in the fireball as it decelerates, using eqn (\ref{eq-c2-08}). When $\Gamma \rightarrow 1$, we shift to the non-relativistic expression given by eqn (\ref{eq-c2-11}).

\bibliography{references}
\end{document}